\documentclass[acmsmall]{acmart}

\usepackage[T1]{fontenc}

\usepackage{amsmath,amssymb,amsfonts}
\usepackage{algorithmic}
\usepackage{natbib}
\usepackage[T1]{fontenc}
\usepackage{graphicx}
\usepackage{subfigure}
\usepackage{array}
\usepackage{textcomp}
\usepackage{xcolor}
\usepackage{stfloats}
\usepackage{mathrsfs}
\usepackage{amsthm}
\usepackage{multirow}
\usepackage{inputenc}
\usepackage[normalem]{ulem}
\useunder{\uline}{\ul}{}
\usepackage{longtable}
\usepackage{booktabs}
\usepackage{pifont}
\usepackage{tabu}
\usepackage{supertabular}
\usepackage{lscape}
\usepackage{threeparttable}
\usepackage{booktabs}

\theoremstyle{definition}

\def\BibTeX{{\rm B\kern-.05em{\sc i\kern-.025em b}\kern-.08em
    T\kern-.1667em\lower.7ex\hbox{E}\kern-.125emX}}

\makeatletter

\newcommand{\Rmnum}[1]{\expandafter\@slowromancap\romannumeral #1@}
\makeatother

\AtBeginDocument{%
  \providecommand\BibTeX{{%
    \normalfont B\kern-0.5em{\scshape i\kern-0.25em b}\kern-0.8em\TeX}}}

\setcopyright{acmcopyright}
\copyrightyear{2018}
\acmYear{2018}
\acmDOI{XXXXXXX.XXXXXXX}

\acmJournal{JACM}
\acmVolume{37}
\acmNumber{4}
\acmArticle{111}
\acmMonth{8}

\begin{document}

\title{Exploring Blockchain Technology through a Modular Lens: A Survey}


\author{Minghui Xu}
\affiliation{%
  \institution{Shandong University}
  \city{Qingdao}
  \country{China}}
\email{mhxu@sdu.edu.cn}

\author{Yihao Guo}
\affiliation{%
  \institution{Shandong University}
  \city{Qingdao}
  \country{China}}
\email{yhguo@mail.sdu.edu.cn}

\author{Chunchi Liu}
\affiliation{%
  \institution{Ernst \& Young}
  \city{Shanghai}
  \country{China}}
\email{Peter.CC.Liu@cn.ey.com}

\author{Qin Hu}
\affiliation{%
  \institution{Indiana University-Purdue University Indianapolis}
  \city{Indianapolis}
  \country{USA}}
\email{Qinhu@iu.edu}

\author{Dongxiao Yu}
\affiliation{%
  \institution{Shandong University}
  \city{Qingdao}
  \country{China}}
\email{dxyu@sdu.edu.cn}

\author{Zehui Xiong}
\affiliation{%
  \institution{Singapore University of Technology and Design}
  \city{Singapore}
  \country{Singapore}}
\email{zehui_xiong@sutd.edu.sg}

\author{Dusit Niyato}
\affiliation{%
  \institution{Nanyang Technological University}
  \city{Singapore}
  \country{Singapore}}
\email{DNIYATO@ntu.edu.sg}

\author{Xiuzhen Cheng}
\affiliation{%
  \institution{Shandong University}
  \city{Qingdao}
  \country{China}}
\email{xzcheng@sdu.edu.cn}

\renewcommand{\shortauthors}{Xu et al.}

\begin{abstract}
Blockchain has attracted significant attention in recent years due to its potential to revolutionize various industries by providing trustlessness. To comprehensively examine blockchain systems, this article presents both a macro-level overview on the most popular blockchain systems, and a micro-level analysis on a general blockchain framework and its crucial components. The macro-level exploration provides a big picture on the endeavors made by blockchain professionals over the years to enhance the blockchain performance while the micro-level investigation details the blockchain building blocks for deep technology comprehension. More specifically, this article introduces a general modular blockchain analytic framework that decomposes a blockchain system into interacting modules and then examines the major modules to cover the essential blockchain components of network, consensus, and distributed ledger at the micro-level. The framework as well as the modular analysis jointly build a foundation for designing scalable, flexible, and application-adaptive blockchains that can meet diverse requirements. Additionally, this article explores popular technologies that can be integrated with blockchain to expand functionality and highlights major challenges. Such a study provides critical insights to overcome the obstacles in designing novel blockchain systems and facilitates the further development of blockchain as a digital infrastructure to service new applications.
\end{abstract}




\maketitle

\section{Introduction and Related Work}
\label{sec:introduction}

\subsection{Introduction}
Blockchain has been receiving considerable attention from both academia and industry since the creation of Bitcoin in 2008~\cite{Bitcoin}. In recent years, there has been a great amount of effort to improve its performance in terms of scalability and security. Scalability enhancement is mainly done by layer-1 solutions such as committee-based consensus protocols and sharding, and layer-2 approaches such as off-chain channels and cross-chain mechanisms. Security analysis has been employed to theoretically prove the security properties of classic blockchain protocols including Bitcoin and Ethereum, or promote trust among users in a few well-respected permissionless blockchains. At the application layer, cryptocurrency and non-fungible token (NFT) are two key areas of focus. According to Garnter's hypercurve, new applications such as Web 3.0 and Metaverse, which take blockchain as one of the key functional primitives, are under development and the related ``killer apps'' are still in their early stages. While there are over 10,000 permissionless blockchains in existence, most of them are adopted for cryptocurrencies. Applying permissioned blockchains in traditional industries is still under exploration. There is not enough information to evaluate whether or not we have benefited from permissioned blockchains, as most applications are inaccessible to the public.

Even though there have been significant improvements in blockchain performance, several grand challenges still exist. First, how to further improve scalability since more widely-used applications have been emerging and require more scalable blockchain protocols. Prism~\cite{bagaria2019prism} demonstrates the possibility of estimating the physical limits of the Proof-of-Work (PoW) protocol~\cite{gervais2016security}. This raises the following question: what is the upper limit of blockchains' performance?
Second, how to ensure interoperability among multiple heterogeneous blockchains, particularly in the absence of trusted third parties. In the past 15 years, a large number of blockchains have been developed to meet a variety of application needs, but they lack a unified standard, leading to strong heterogeneity. Hence there is a great demand on blockchains to be highly flexible in order to meet diverse needs and also support cross-chain interoperability. 
Finally, popular blockchain applications are rapidly evolving. Permissionless blockchains are expanding to include uses such as NFTs, Web 3.0, and the Metaverse, while permissioned blockchains are focusing more on specific areas. In addition, blockchain is being integrated with other technologies to create  information systems such as data sharing platforms and access control modules. These developments drive us to reconsider how 
blockchains can be beneficial in developing new applications and techniques in future. In this article, we review both classics and status quo of blockchain to respond to the above challenges.

\subsection{Related Work}

We briefly summarize relevant survey articles and literature reviews regarding blockchain frameworks, consensus protocols, incentive mechanisms and applications. A comparison study is reported in Table~\ref{tb:survey:comparison}. 

Layered architectures have been widely adopted to analyze blockchain systems. Hileman and Rauchs provided a benchmarking study~\cite{hileman2017global} to specifically decouple a blockchain system into three layers: protocol, network, and application. Gao \textit{et el.}~\cite{gao2018survey} proposed a different three-layered blockchain framework consisting of a network layer, a data layer, and an application layer. Wang \textit{et el.}~\cite{wang2019survey} presented an implementation stack comprised of network, data, and application layers. 
Besides, some surveys provide new perspectives on understanding and analyzing blockchain systems. The software architecture of cryptocurrency, smart contract applications, and reputation systems were briefly discussed in ~\cite{aldweesh2016survey}. Bartoletti \textit{et el.}~\cite{bartoletti2017general} proposed a general-purpose framework to integrate the data derived from Bitcoin and Ethereum by focusing on the interconnection between data organization protocols and network protocols. 
Ballandies \textit{et al.}~\cite{ballandies2018decrypting} provided a conceptual architecture containing four components and 19 attributes to depict blockchains and Cryptoeconomic Design (CED). 
A few frameworks were used to analyze research interests. Tavares \textit{et el.}~\cite{tavares2018survey} counted papers and citations concerning blockchain research. Risius and Spohrer~\cite{risius2017blockchain} concluded that blockchain research predominantly focuses on technical design, while less on application, governance, and value creation. 

In~\cite{wang2019survey}, the authors provided a valuable resource for those interested in the design and implementation of distributed consensus systems and incentive mechanisms. 
Xiao \textit{et el.}~\cite{xiao2019survey} conducted a comprehensive survey on consensus algorithms under an analytic framework that is comprised of five components: block proposal, block validation, information propagation, block finalization, and incentive mechanism. Five different consensus algorithms were reviewed by Du \textit{et el.}~\cite{mingxiao2017review}. Zheng \textit{et el.}~\cite{zheng2017overview} analyzed six popular consensus algorithms from the perspectives of identity management, energy consumption, and fault tolerance. They also mentioned five future directions worthy of being noted. Dinh \textit{et al.}~\cite{dinh2018untangling} designed a benchmarking framework named BLOCKBENCH to untangle blockchains from the perspective of data processing. 

In the view of incentive mechanisms, Liu \textit{et el.}~\cite{liu2019survey} conducted a comprehensive survey on the game-theoretical approaches adopted by blockchains. In particular, they discussed how game theory can mitigate security issues such as selfish mining attacks, majority attacks, and DoS attacks. Besides, game theory can assist mining management through proper computational power distribution, reward mechanism, and block size selection. Some incentive schemes adopted by blockchain were briefly analyzed in~\cite{yu2018survey}.

PoW-based cryptocurrencies were specifically surveyed in~\cite{mukhopadhyay2016brief} considering the strengths and weaknesses of their mining strategies. Tschorsch and Scheuermann~\cite{tschorsch2016bitcoin} provided a detailed survey on cryptocurrencies from a holistic technical perspective. In particular, these two authors started with an in-depth analysis on Bitcoin basics and its security and privacy, then discussed a broader field of consensus algorithms and the blockchain networks.
Market targeting and usage, business models, and maturity of blockchain systems were discussed in~\cite{hileman2017global}. Emerging applications of blockchain such as IoT, big data, and cloud computing were covered by~\cite{gao2018survey}. The literature review by Casino \textit{et el.}~\cite{casino2018systematic} presents a review and classification of ten various blockchain-based applications. 


\begin{table}[!htbp]
\caption{Comparison of Recent Blockchain System Surveys}
\begin{center}
\begin{tabular}{m{3.5cm}<\centering m{2.5cm}<\centering m{1.5cm}<\centering }
\toprule[1pt]
\textbf{Perspectives of blockchain System Surveys} & \textbf{Recent papers} & \textbf{Covered in this survey} \\
\midrule[0.8pt]
Blockchain famework and architecture                    & ~\cite{hileman2017global,tavares2018survey,aldweesh2016survey,bartoletti2017general,gao2018survey,wang2019survey,zheng2017overview,risius2017blockchain,lu2018blockchain,dinh2018untangling}  & \checkmark                       \\ 
\hline
Consensus protocol                          & ~\cite{mukhopadhyay2016brief,mingxiao2017review,yu2018survey,xiao2019survey,wang2019survey,tschorsch2016bitcoin,zheng2017overview,zheng2018blockchain,chalaemwongwan2018state,sankar2017survey,dinh2018untangling,zhang2018towards}                                      &  \checkmark \\
\hline
Game theory and incentive mechanism          &  ~\cite{yu2018survey,liu2019survey}                                               &\checkmark                        \\ 
\hline
Distributed ledger structure             &-                                                             & \checkmark \\ \hline
Application and business model              & ~\cite{hileman2017global,tavares2018survey,gao2018survey,ballandies2018decrypting,casino2018systematic,zheng2018blockchain,lu2018blockchain,dinh2018untangling}                           & \checkmark                  \\ 
\bottomrule[1pt]
\end{tabular}
\end{center}
\label{tb:survey:comparison}
\end{table}

\subsection{Contributions}

We summarize our contributions as follows.
\begin{enumerate}
\item Comprehensive overview: In contrast to recent surveys that focus on specific components of blockchain technologies, this article presents both a macro- and a micro-level analysis on blockchains systems. The micro-level investigation offers an extensive overview on the crucial blockchain components, including network, consensus, and ledger structure,  which provides a complete and thorough comprehension of blockchain technologies. 
\item Modular framework: We introduce a modular framework to facilitate the deep understanding of blockchain technologies, through the use of a modular blockchain analytic framework, which decomposes a blockchain system into interacting modules, making it easier to comprehensively analyze each component. This framework can facilitate the future design of scalable, flexible and application-adaptive blockchain systems that can meet the unique requirements of diverse applications. 
\item Future directions and challenges: Furthermore, we offer insights into the potential future directions on blockchain applications and highlight the open research challenges. 
While blockchain is getting an increasing popularity, it still has critical obstacles that need to be addressed, which includes scalability, security, modular design, storage, privacy, and the integration with AI. This article presents a critical review on these challenges involved in adopting blockchain  as an infrastructure, ultimately paving the way towards the successful upgrading and updating of blockchain systems.
\end{enumerate}

\subsection{Paper Organization}
The rest of the article is structured as follows. Section~\ref{sec:blockchain:basics} presents the blockchain basics to warm up. In Section~\ref{sec:scaling}, we first illustrate the evolution of scaling techniques to introduce the popular blockchain systems from a macro perspective, then propose the modular blockchain analytic framework for the purpose of guiding the following micro-level studies on the key blockchain building blocks. Particularly, we elaborate on the three essential components of the framework, namely blockchain network, consensus, and distributed ledger, in Sections~\ref{architecture}–\ref{ledger}. Section~\ref{sec:techniques} highlights the relevant techniques that can be integrated with blockchain to expand its functionality and presents promising application scenarios. In Section~\ref{sec:challenge}, we report the open challenges and future directions, leading to conclude this article in Section~\ref{conclusion}.

\section{Blockchain Basics}
\label{sec:blockchain:basics}

In general, Distributed Ledger Technology (DLT) refers to sharing, replicating, and synchronizing a digital ledger across a network in which there exists no trusted authority. Blockchain is a type of DLT that organizes a digital ledger as a chain of blocks linked by hashes. This section provides an overview on the basics of blockchain, including network models, structures, types, features, account models, and design philosophies.

\subsection{Network Model}

\begin{figure}[htbp]
\centering
\centerline{\includegraphics[width=0.6\textwidth]{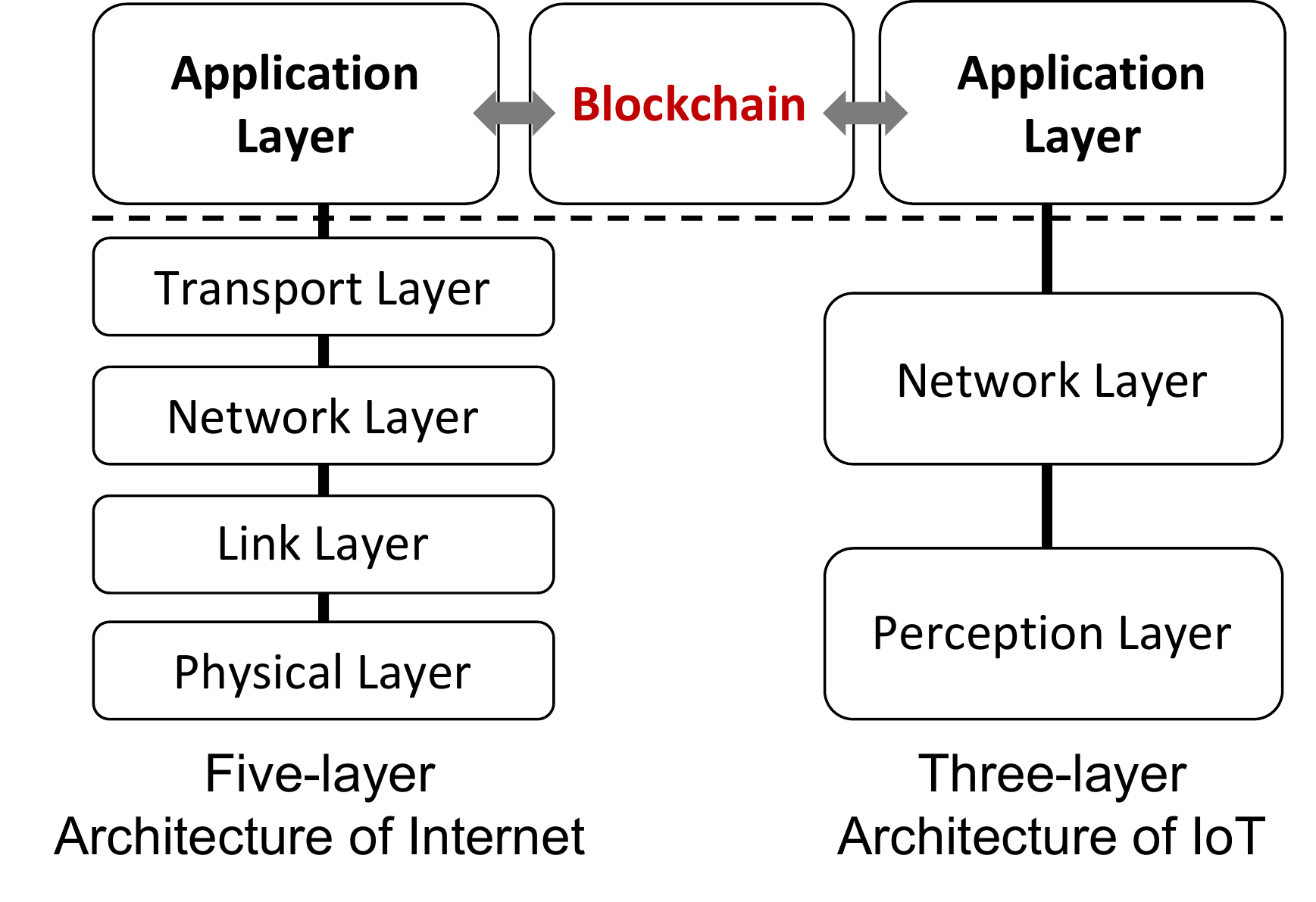}}
\caption{Blockchain and the architectures of the Internet and IoT.}
\label{architecture:internet}
\end{figure}

Fig.~\ref{architecture:internet} demonstrates the role blockchain plays from the perspective of network architecture considering both the Internet of Things (IoT) and the Internet. The Internet takes the most commonly used five-layer model (i.e., physical layer, link layer, network layer, transport layer, and application layer) while IoT is usually structured in three layers (perception layer, network layer, and application layer). One can see that blockchain resides on the application layer in both models. This implies that blockchain currently does not get sufficient involvement in the underlying layers, which greatly hinders it from fully releasing its potential and hence results in many problems such as high latency and privacy leak. Therefore in future developments, blockchain needs to be gradually integrated with the underlying layers in order to become a fully-fledged infrastructure. To achieve this, we propose a modular blockchain analytic framework in this article that allows for scrutinizing  a blockchain system from a modular perspective. This framework can facilitate the design of new blockchain systems as well as blockchain-based information infrastructures.

\begin{figure*}[htbp]
\centering
\centerline{\includegraphics[width=0.95\textwidth]{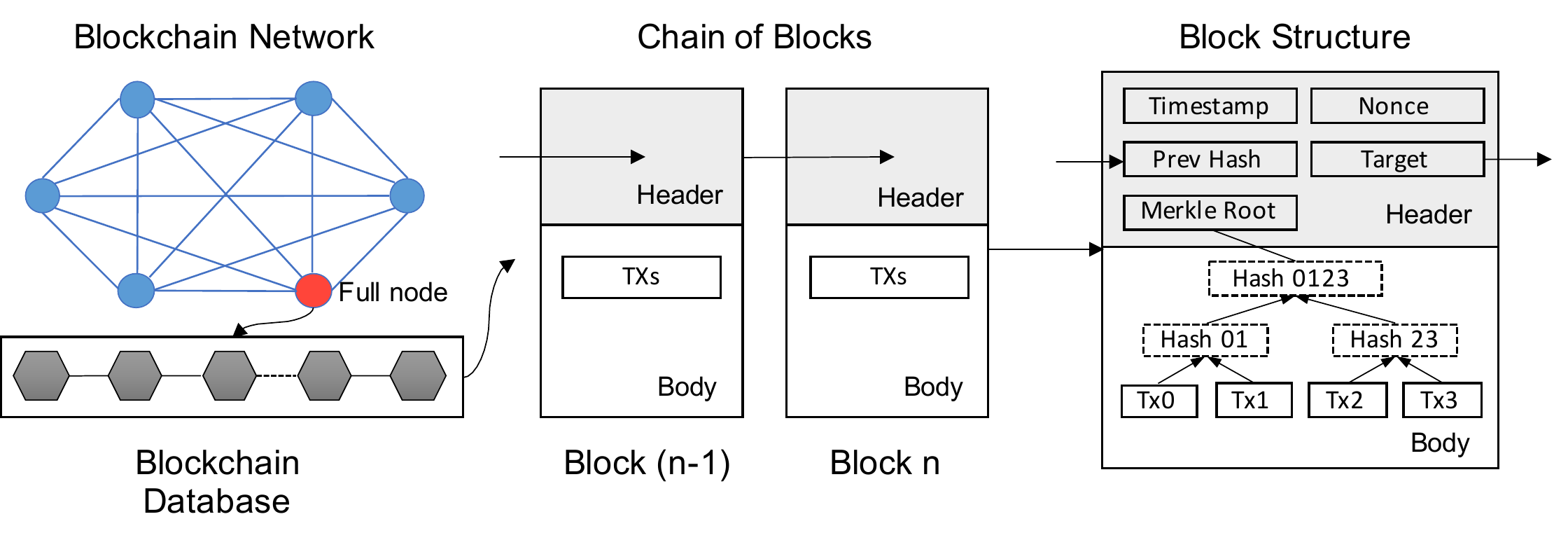}}
\caption{Blockchain structure: blockchain database, chain of blocks, and block structure.}
\label{blockchain:structure}
\end{figure*}

A blockchain is a chain of blocks maintained by multiple nodes in a peer-to-peer network, as shown in Figure~\ref{blockchain:structure}. 
Fundamentally, there are three types of nodes: full nodes, light nodes, and validators (or miners). Each \emph{Full} node maintains a complete copy of the ledger, validates all transactions and blocks, and participates in the network by relaying information for others. In contrast, a \emph{light} node only stores the block headers, making it faster for verification but more vulnerable to security risks such as double spending attacks and chain forks. Note that light nodes rely on full nodes to provide them with information on demand. They are useful for lightweight devices with limited storage and computational power. A \emph{Validator} (or a \emph{miner}) is a full node that participates in the consensus protocol to create new blocks. In this article, we focus on analyzing the ``kernel'' of a blockchain, which mainly constitutes of consensus nodes.

\subsection{The Basic Structure}

Each block, except for the genesis block, contains a header and a body. The body typically includes serialized transactions, while the header consists of a timestamp, a nonce, the previous hash, the block hash, and the Merkle root. A block links to its previous block by storing the latter's hash, i.e., the previous hash. The Merkle root, as an accumulator, is used for set-membership tests. To create a Merkle tree, the transactions in a block body are input as leaves, and each parent node is the hash of the values of its two children. With a Merkle tree, a block only needs to include the Merkle root as the fingerprint of the entire set of transactions stored in its body. Each leaf in a Merkle tree can be verified through a Merkle path from the root, which is of size $O(\log N)|H|$, where $N$ is the number of transactions in the block and $|H|$ is the hash length.

\subsection{Blockchain Type}
Typically, there exist three types of blockchains based on scale and openness: public, private, and consortium. Public blockchains are also termed as permissionless blockchains, while private and consortium blockchains are permissioned. 

\begin{itemize}
\item \textit{Public Blockchains:} 
Public blockchains, with Bitcoin and Ethereum as two typical examples, are decentralized, open to the public, and self-governed. Anyone can join or leave a public blockchain at any time, and all can participate in the consensus process as validators. Public blockchains often use incentives, such as transaction fees and mining rewards, to encourage nodes to maintain the normal operations of the network. However, they face challenges from malicious nodes that can launch token stealing~\cite{atzei2017survey}, selfish mining~\cite{sapirshtein2016optimal, eyal2018majority}, Sybil~\cite{douceur2002sybil}, eclipse~\cite{singh2006eclipse}, and various other attacks. Additionally, as distributed ledgers can be publicly accessed, concerns about privacy and anonymity are duly noted~\cite{li2017survey}. Power consumption is another significant issue in public proof-of-work-based blockchains, especially when they are deployed in a large scale.

\item \textit{Private Blockchains:} 
Private blockchains, with Multichain~\cite{multichain}, Monax\cite{monax}, 
and Blockstack~\cite{Blockstackapp} as typical examples, are used and governed by single organizations. Instead of being open to anyone, the access to a private blockchain is restricted and requires a verified invitation. As a result, private blockchains are not decentralized, but permission-based and closed, which makes it easier to manage them and provides better privacy, but sacrifices decentralization and openness. The operator of a private blockchain has the ability to override, roll back, delete, and edit blocks, which undermines blockchain's trustless property. Private blockchains are also generally smaller in scale, which can make them more efficient and secure. Many companies prefer to use a private blockchain because it allows them to quickly identify the source of errors and minimize losses.

\item \textit{Consortium Blockchains:} 
Consortium blockchains are permissioned and partially decentralized, falling between public and private ones. Examples include Hyperledger~\cite{Hyperledger}, Quorum\cite{Quorum}, Ripple~\cite{Ripple}, and Corda~\cite{brown2016corda}. Unlike private blockchains, consortium blockchains are not owned by a single entity but by multiple organizations. They are well-suited for building trustless applications among different organizations that prioritize privacy and want to take advantage of the blockchain properties.
\end{itemize}

\subsection{Blockchain Feature}
\label{blockchain:features}
Blockchain technologies have been applied to a wide range of fields, including finance, supply chain management, unmanned aerial vehicle swarms, and cloud services. These applications rely on blockchain's key properties of 
immutability, fault tolerance, and openness. 

\begin{itemize}

\item \textit{Immutability:} 
Immutability refers to the fact that blockchains are append-only, and once a block has been accepted into a blockchain network, it cannot be modified. This is achieved through the use of a consensus protocol to reach agreement on the next authenticated state of the blockchain, as well as the exploitation of hashes, digital signatures and other cryptographic techniques to ensure the integrity of the data. Immutability helps to prevent disputes, preserve data integrity, and facilitate auditing, but it does not allow private and sensitive information to be removed once publicly recorded in a blockchain.

\item \textit{Fault Tolerance:} 
Fault tolerance is the ability of a system to continue operating correctly even when some of its components fail or malfunction. In decentralized networks, fault tolerance is achieved through the use of redundant components and the distribution of functions across multiple nodes. This allows the system to continue functioning even if some nodes fail or are unavailable. Blockchain provides strong fault tolerance by avoiding reliance on a single point of control and eliminating the need for a central authority manipulating the network. Public blockchains are designed to be as decentralized as possible, with the ledger being maintained through replication across many distributed nodes. Private and consortium blockchains, on the other hand, sacrifice some level of fault tolerance in exchange for increased security.

\item \textit{Openness (public blockchains):} 
Openness in a blockchain means that the blockchain network is fully accessible to the public, allowing anyone to join and leave at its will, send transactions, view the ledger, and participate as a validator in the consensus process. This is in contrast to traditional banks and service providers, which do not grant users accesses to their complete ledgers. In addition, many blockchain implementations are open source, enhancing trust on them and accelerating blockchain developments. Openness makes blockchains transparent and auditable; it also enables them to scale more easily.
\end{itemize}

\subsection{UTXO and Account/Balance}

There are mainly two models to represent transactions and balances in a blockchain: the unspent transaction output (UTXO) model and the account/balance model. In the UTXO model, each transaction spends money from UTXOs received in previous transactions and creates new UTXOs that can be spent in future. To be verified, a legitimate UTXO must specify its owner and the amount of money it represents. The account/balance model is similar to that used by a traditional bank, where each account is associated with a balance and transactions can transfer money from one account to another. Both models have their own advantages and disadvantages in terms of security, scalability, and simplicity.

\begin{itemize}
\item \textit{Security:} In the UTXO model, a valid transaction must include valid UTXOs whose total value is strictly greater than the total amount of the outputs. This ensures that the transaction is not creating new cryptocurrency out of thin air and that the outputs can be covered by the inputs. Because each UTXO can only be spent once, double-spending attacks can be prevented. In the account/balance model, a simple method to prevent double-spending attacks is to include a counter in transactions. If there are conflicting transactions, validators choose the one with the lowest counter value.
\item \textit{Scalibility:} The UTXO model supports the parallel processing of transactions because UTXOs are atomic and fragmented. Hence multiple transactions from the same user can be paid with different UTXOs in parallel. In contrast, transactions must be processed serially with the account/balance model.
\item \textit{Simplicity:} The UTXO model is more complex because the balance must be calculated from UTXOs, and a transaction must include all the UTXOs it is spending. In contrast, a transaction in the account/balance model only needs to verify that the sending account has sufficient fund to cover the payment. This makes transaction validation simpler and easier to implement. Additionally, the simplicity of the account/balance model makes it more straightforward to implement complex application logics such as those in Decentralized Applications (DApps).
\end{itemize}

\subsection{Blockchain Philosophy}

The essence behind blockchain lies in two aspects: the generation of honesty (related to node registration and consensus processes), and the preservation of honesty (related to the ledger structure and the distributed storage).

{\bf Generation of Honesty.}
Let $\mathcal{G}$ be a set of blockchain nodes. We can define the Honesty Generation Function ($HGF$) to generate a valid block $\mathcal{B}$ among $\mathcal{G}$, denoted by $\mathcal{B} = HGF(\mathcal{G})$. Due to nodes' faulty behaviors and network asynchronization, the probability that $\mathcal{B} = HGF(\mathcal{G})$ is not guaranteed to be 1 but instead a probability $P<1$. Blockchain protocols aim to provide $HGF$ with a high $P$ and low communication/time complexity. 

There are two factors that impact $P$: $\mathcal{G}$ and $HGF$. $\mathcal{G}$ depends on the registration policy of the blockchain. Permissioned blockchains tend to have a higher percentage of honest nodes, while it is harder for permissionless blockchains to guarantee high-quality participants. 
The $HGF$ largely depends on the blockchain consensus protocol. Simple designs are to set $HGF$ as a random sampling in that $P$ is entirely based on $\mathcal{G}$, which demands $\mathcal{G}$ to fulfill the honest majority requirement. For example, if we have 50\% faulty nodes, the probability $P$ based on random sampling is 50\%. Sophisticated designs of $HGF$, on the other hand, can help promote $P$ or loosen the requirement on $\mathcal{G}$. For example, a consensus algorithm may choose a trustworthy leader to propose a block, which can increase the probability of confirming an honest block.

{\bf Preservation of Honesty.}
After generating honesty, it is necessary to preserve it, both temporally and spacially. Temporal security emphasizes the temporal order of the transactions in the ledger $R$, i.e., $R$ should correctly record the transactions according to their occurrence time. The ground-truth record after the $t_i$th transaction is $R_{i+1} = g_t(t_1, t_2, \dots, t_i)$, where $g_t(\cdot)$ denotes a temporally-secured distributed transaction processing function. Or we can rewrite $R$ in a recurrence form: $R_{i+1} = g_t(R_i, t_i)$. A transaction cannot timely surpass one that has happened earlier, which also provides fairness. 
For the verification procedure, one should efficiently verify the sequential correctness of $R$. 

However, one replica of the ledger alone, which can be modified into a wrong one, cannot guarantee the security of a blockchain. So blockchains require spacial security to protect ledger $R$. A blockchain runs $g_s(\cdot)$, a storage distribution function, to generate $r_1, r_2, \dots, r_n$, where $r_i$ is stored in the $i$th node. Reversely, $g_s^{-1}(\cdot)$ is the inverse function  used to recover $R = g_s^{-1}(r_1, r_2, \dots, r_n)$. 
A naive design may have $r_1=r_2=\dots=r_n = g_s(R)$, which has been adopted by the current mainstream blockchain systems, but it suffers from the drawbacks of low efficiency and high storage overhead. Sophisticated designs can make $g_s(\cdot)$ elaborate and $r_1, r_2, \dots, r_n$ storage-space-friendly (e.g., by erasure coding~\cite{qi2020bft}), which is a promising approach for the novel future ledger designs.

\section{An overview on Blockchain Systems from Macro and Micro Perspectives}
\label{sec:scaling}

In this section, we start with a comprehensive examination on popular blockchain systems from a \textit{macroscopic} perspective, focusing on their scalability and security. Specifically, we classify them into two categories, i.e., layer 1 and layer 2, which are respectively detailed in sections \ref{sec:Layer-1} and \ref{sec:Layer-2}. Then we introduce the Modular Blockchain Analytic Framework in section \ref{sec:framework} as a microscopic lens to dissect and deeply analyze these blockchain systems. This framework complements our macroscopic analysis and enables us to explore the underlying building blocks (i.e., blockchain network, consensus protocol, and distributed ledger) in more detail.

Blockchain is a promising technology for decentralized applications, but it suffers from several limitations that hinder its scalability and security. Two closely-related and important issues are the low transaction rate and high transaction processing latency, which make it difficult for blockchain systems to handle a large amount of data. To address these issues, a class of solutions known as Layer-1 scalability were developed, which aim to incrementally improve the blockchain performance by adding or modifying fundamental attributes such as consensus algorithm, block size, and network architecture.
Besides that, a completely different class of works, called Layer-2 protocols, emerged, which were built on top of the existing blockchain networks to help increase the capacity of the underlying blockchain without modifying its fundamental attributes. In the following two subsections, we elaborate on the layer-1 and layer-2 solutions considering properties such as scalability, security, and efficiency.

\subsection{Layer 1} \label{sec:Layer-1}
\begin{figure*}[htbp]
\centering
\centerline{\includegraphics[width=1\textwidth]{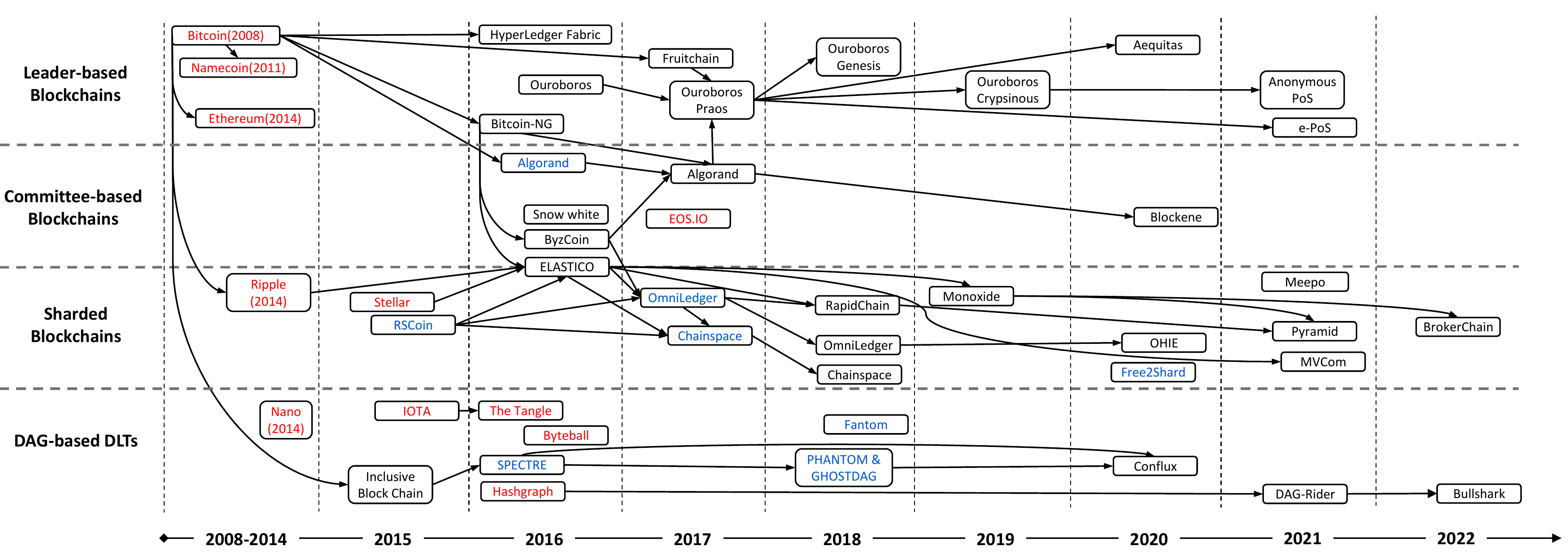}}
\caption{Layer-1 techniques of blockchain systems organized in the chronological order. Black, \textcolor{blue}{blue}, and \textcolor{red}{red} fonts respectively refer to peer-reviewed papers, \textcolor{blue}{preprints}, and \textcolor{red}{whitepapers}.}
\label{Timeline}
\end{figure*}

To provide an overview on the evolution of blockchain systems and set a stage for the rest of the article, we study the popular blockchain systems and organize them in a chronological order, as shown in Figure~\ref{Timeline}. The edges in the figure are directed, meaning that an edge pointing from blockchain $A$ to blockchain $B$ indicates that $B$ solves certain problems of or improves upon $A$. Such relationships between blockchain systems have been derived from their peer-reviewed papers, preprints, and white papers. While some works may have appeared in multiple locations or have multiple versions, Figure~\ref{Timeline} only includes representative ones for clarity. We divide Layer-1 solutions into four categories: leader-based, committee-based, sharded, and DAG-based.

\subsubsection{Leader-based Blockchains}
\label{leader:based}
A leader-based blockchain directly selects a leader to propose the next block without forming a committee or partitioning the nodes in advance. In this category, blockchain development starts from the earliest ancestor Bitcoin to improve scalability and optimize resource consumption. 

The emergence of Bitcoin and Ethereum laid the foundation for the development of other blockchain platforms. Bitcoin, which was introduced in 2008, employs a proof-of-work consensus mechanism in which miners compete for computational power and the winner is selected as the leader for a particular time slot. 
In 2011, Namecoin was introduced as the first alternative cryptocurrency (altcoin), enabling decentralized naming services and providing a merged mining consensus algorithm. In 2014, Ethereum was introduced as a blockchain-based distributed computing platform that includes the (theoretically) Turing-complete Ethereum Virtual Machine (EVM), which enables the creation of smart contracts and expands the capabilities of blockchain beyond payment. 
Ethereum takes a modified proof-of-work consensus mechanism called Ethash, which is resistant to abnormal mining activities involving specialized hardware such as application-specific integrated circuits (ASICs), graphics processing units (GPUs), and field-programmable gate arrays (FPGAs). 
The Hyperledger project~\cite{Hyperledger}, which was started by the Linux Foundation in December 2015, includes Hyperledger Fabric~\cite{hyperledgerfabric2018}, a permissioned blockchain system that supports configurable consensus algorithms and access mechanisms.


Bitcoin-NG~\cite{Bitcoin-NG} and Fruitchain~\cite{pass2017fruitchains} were both designed to improve the scalability of Bitcoin by modifying its ledger structure. Like Bitcoin, Bitcoin-NG makes use of a proof-of-work consensus algorithm to elect a leader. The leader is responsible for serializing transactions and constantly proposing microblocks at a predefined rate until a new leader takes over and starts a new epoch. With a similar idea, Fruitchain attaches multiple microblocks (called ``fruits'') to key blocks.

One concern about proof-of-work (PoW) based blockchains, particularly Bitcoin, is their high energy consumption. It was estimated that Bitcoin consumed more than 26,000 MW of power in November 2019~\cite{Realtimebitcoin}, which is equivalent to the total output of 26 nuclear power plants (with each typically having an output of 1000 MW)\footnote{Estimated by https://en.wikipedia.org/wiki/List\_of\_nuclear\_power\_stations}. To address this issue, the concept of proof-of-stake (PoS)~\cite{originalPoS} was introduced in the Bitcoin forum. PoS relies on users' stake rather than computational power to select a leader. Ethereum is striving to transition from PoW to PoS, which means that it intends to bid farewell to the era of large-scale mining and reduce energy consumption by 99.99\% in estimation. 

Snow White and Ouroboros are leader-based blockchains with provably secure PoS-based consensus protocols. Ouroboros~\cite{Ouroboros} realizes a randomness beacon achieved by the \textit{G.O.D Coin tossing} algorithm that makes use of a coin-tossing game and verifiable secret sharing~\cite{feldman1987practical}. The research group behind Ouroboros (including IOHK~\cite{IOHK}, the University of Edinburgh, etc.) also built the Ouroboros family that includes the following leader-based designs: Ouroboros (provably secure), Ouroboros Praos (adaptively secure)~\cite{david2018ouroboros}, Ouroboros Genesis (dynamically available)~\cite{badertscher2018ouroboros}, and Ouroboros Crypsinous (privacy-preserving)~\cite{kerber2019ouroboros}. 
CloudChain~\cite{CloudChain} presents a leader-based consensus algorithm based on a shared-memory model where nodes communicate synchronously by direct memory accesses within a cloud.

Since 2020, leader-based blockchains with new features such as security and fairness have been constructed~\cite{kelkar2020order, saad2021pos}. 
Aequitas~\cite{kelkar2020order} achieves order-fairness to solve the problem of incorrect final transaction ordering caused by malicious leaders. Saad \textit{et al.}~\cite{saad2021pos} proposed the e-PoS algorithm to resist the centralization of PoS-based blockchain networks. 

\subsubsection{Committee-based blockchains}
\label{commttee:based}
A committee aims to improve scalability by selecting a small group of validators to participate in a consensus process. This reduces the communication cost when executing a consensus protocol. Considering that running a Practical Byzantine Fault Tolerance (PBFT) consensus algorithm involves a message complexity of $O(N^2)$ without a view change, running the PBFT on $2N$ nodes is four times slower than running it on $N$ nodes. A committee must be elected randomly to avoid adversaries from predicting and attacking the committee members. To achieve randomness, Algorand~\cite{Algorand} employs a sortition algorithm that leverages verifiable random functions. Blockene, proposed by Satija \textit{et al.}~\cite{satija2020blockene}, is similar to Algorand but discloses the identities of committee members before they participate in the consensus protocol. ELASTICO~\cite{elastico} designates a final committee made up of members who are randomly selected from each shard. ByzCoin~\cite{ByzCoin} adopts a dynamic committee formed by a sliding-window mechanism, which dynamically reorganizes the committee to include new block miners and remove old ones. Snow White~\cite{bentov2016snow} presents a provably secure proof-of-stake (PoS) based blockchain protocol that uses epoch-based committee selection to ensure that all nodes have the same view of the updated committee. EOS.IO~\cite{EOS} forms a committee in each round by selecting 21 producers, who then run the delegated proof of stake (DPOS) consensus algorithm.

\subsubsection{Sharded blockchains}
\label{sharded}
Sharding refers to partitioning\footnote{The definition of sharding is different from partitioning in the database research community. Database sharding is a specific type of partitioning, namely horizontal or vertical partitioning~\cite{corbett2013spanner}.} a blockchain network to enable faster transaction processing and better scalability. Generally speaking, a blockchain network can be sharded into smaller sub-networks. 
Ripple~\cite{Ripple}, Stellar~\cite{Stellar}, RSCoin~\cite{danezis2015centrally}, Meepo~\cite{zheng2021meepo}, Pyramid~\cite{hong2021pyramid}, and MVCom~\cite{huang2021mvcom} are examples that implement sharding protocols.

Ripple~\cite{Ripple} is one of the earliest sharded blockchains. Particularly, each Ripple node maintains a Unique Node List (UNL) recording a set of other nodes to interact with. Nodes are divided into several shards and all nodes within a shard are connected to each other. Each shard executes the Ripple Protocol Consensus Algorithm (RPCA). In Stellar~\cite{Stellar}, users are partitioned into overlapping shards (called quorum slices). Stellar adopts a type of Federated Byzantine Agreement (FBA), also referred to as the Stellar Consensus Protocol (SCP). RSCoin~\cite{danezis2015centrally} was proposed in December 2015 as a Centrally Banked Cryptocurrency, which improves scalability at the cost of decentralization. The central bank delegates the authority of validating transactions to different shards controlled by banks and other institutions. Zheng \textit{et al.}~\cite{zheng2021meepo} proposed Meepo, a sharded scheme designed for consortium blockchains, which aims to enhance cross-shard efficiency based on methods named cross-epochs and cross-calls. 
Pyramid, presented in~\cite{hong2021pyramid}, proposes a sharding approach that involves layers. Unlike Rapidchain~\cite{Rapidchain} and Monoxide~\cite{Monoxide}, which require dividing cross-shard transactions into multiple sub-transactions, Pyramid does not necessitate this; instead, its solution entails certain shards storing the states of other shards to construct a layered structure, which is capable of efficiently processing cross-shard transactions.
According to~\cite{huang2021mvcom}, MVCom employs an algorithm for online stochastic exploration to arrange the most dependable committee, leading to an improvement in performance.

Sharding can improve the blockchain scalability at the cost of introducing security risks. Sharding techniques always address these concerns by guaranteeing randomness and atomicity. For example, ELASTICO exploits PoW-based randomness generation to randomly distribute nodes into shards. OmniLedger combines VRF and RandHound~\cite{randomhound} for stable and fair sharding, and employs Atomix to guarantee atomic cross-shard operations. 
OHIE~\cite{yu2020ohie} aims to provide a safe but elegant blockchain system, running multiple parallel instances of the Nakamoto consensus protocol. 
Chainspace~\cite{Chainspace} addresses atomic cross-shard transactions using the S-BAC protocol, which tolerates up to 1/3 Byzantine nodes. 
RapidChain was presented in 2018~\cite{Rapidchain}, which can surpass ElASTICO and OmniLedger in terms of latency and throughput according to the presented experimental studies. 
Monoxide~\cite{Monoxide} improves the blockchain scalability by generating asynchronous shards, called consensus zones, to maintain independent communication, computation, storage, and memory. It introduces eventual atomicity for cross-shard transactions and proposes the use of Chu-ko-nu mining to disperse mining power and reactivate PoW in a sharded blockchain. 
Monoxide is subject to the ``hot-shard'' issue, which was addressed in BrokerChain. BrokerChain~\cite{huang2022brokerchain} is account-based, which achieves fine-grained state partition and account segmentation to improve system throughput. 

\subsubsection{DAG-based Blockchains} \label{dag}
Distributed ledger systems based on Directed Acyclic Graphs (DAGs) are a type of blockchain that expands the conventional linear chain to a two-dimensional graph structure. The ``direct'' and ``acyclic'' respectively indicate a time-based sequence of transactions and the absence of conflicting transactions. 
There are mainly two types of DAG-based blockchains. The first type uses a DAG vertex to represent a transaction. Examples include IOTA~\cite{popov2016tangle}, Byteball~\cite{churyumov2016byteball}, Swirlds Hashgraph~\cite{baird2016hashgraph}, Conflux~\cite{conflux}, and DAG-Rider~\cite{dag-rider}. 
The second type incorporates DAGs into the traditional blockchain model by allowing blocks to have multiple parents, thus improving performance. Examples of this type include the Inclusive Block Chain Protocol~\cite{lewenberg2015inclusive}, SPECTRE~\cite{sompolinsky2016spectre}, Nano~\cite{lemahieu2018nano}, and PHANTOM GHOSTDAG~\cite{sompolinskyphantom}.

IOTA Tangle, proposed in 2015, is a DAG-based blockchain with a ledger structure called ``Tangle''. In Tangle, each vertex corresponds to a transaction, referencing two previous ones. This means that each vertex has to approve two previous transactions that are selected by the Markov chain Monte Carlo (MCMC) algorithm. To maintain consistency in Tangle, pruning is carried out based on the total weights of each path.
Launched in 2016, Byteball (now known as Obyte) presents a unique ledger structure known as the ``storage unit'', which holds essential data such as signature, amount, and other information, similar to the transactions in IOTA. However, unlike Tangle's referencing rules, Byteball incentivizes nodes to reference all earlier vertices using rewards, thus increasing bandwidth usage. 
Hashgraph is a data structure with multiple parallel transaction chains (called ``events'') that are interconnected to each other. Each node manipulates a chain, and the interconnections among chains represent the gossip history among nodes. Recently, Keidar et al.~\cite{dag-rider} introduced DAG-Rider, the first asynchronous Byzantine atomic broadcast protocol with post-quantum security. It has a two-layer structure and can reduce communication complexity compared to Hashgraph while exploiting cryptographic assumptions for increased safety. 
Conflux is a scalable and high-throughput blockchain system that maintains a total order for transactions. Conflux consensus protocol operates on a tree-graph to provide fast transaction confirmation. 

The Inclusive Block Chain Protocol is a pioneer second type DAG-based blockchain, which establishes a formalized blockDAG model. SPECTRE employs a Proof-of-Work (PoW) consensus algorithm and a voting mechanism to determine the order of the blocks in a DAG, thereby guaranteeing consistency.
Nano uses a block-lattice, a type of DAG-based ledger in which each account has its own blockchain to track its transactions and balance history. 
PHANTOM makes further progress towards establishing a robust total order of all transactions by demonstrating that the set of blocks created by honest miners is well-connected. A drawback of PHANTOM is that computing a well-connected set (called the Maximum k-cluster SubDAG) is NP-hard; thus it takes a practical solution using a greedy algorithm called GHOSTDAG.

\subsection{Layer 2} \label{sec:Layer-2}

Layer 2 techniques intend to enhance efficiency, scalability, and interoperability of blockchains by offloading some of the transaction processing and computation tasks from layer 1 while still relying on the security guarantees of the underlying layer 1. We explain two primary layer-2 solutions, namely off-chain and cross-chain techniques. 

\subsubsection{Off-Chain Techniqes} \label{sec:Off-Chain Techniqes}
We divide off-chain techniques into three categories, namely off-chain channels, Blockstack, and Rollup. 

The off-chain channel technology is originated from the lightning network~\cite{poon2016bitcoin}, a system that allows two users who frequently make transactions with each other to create an off-chain payment channel that only updates the balances between them without recording any transaction on the blockchain. This can improve scalability and lower transaction fees.  There are several versions of the lightning network, including LND, c-lightning, and eclair\footnote{https://github.com/bcongdon/awesome-lightning-network}. Besides, there have been several off-chain channel developments, including state channels and virtual channels, which offer different features such as versatility, re-balancing, high capacity, and privacy protection~\cite{dziembowski2018general,dziembowski2019perun, dziembowski2019multi, aumayr2021bitcoin, li2020secure, khalil2017revive, papadis2022payment, green2017bolt, malavolta2017concurrency}. 

Blockstack~\cite{Blockstackapp} offers a layer-2 solution to enhance scalability. It involves the Stacks blockchain (layer 1) and a user-controlled storage system called Gaia (layer 2). Gaia can be hosted on a cloud server or locally, and allows users to create separate storage buckets for different applications. 
It stores data securely with encryption and signatures, while the Stacks blockchain only stores pointers to the data. 
One potential drawback of Blockstack is the storage and processing overhead required for cryptographic operations. However, Blockstack claims that the storage overhead is only 5\% larger than the original file size, which limits the CPU overhead for reads and writes. 

Rollup is a system that aims to improve the efficiency of transactions by compressing multiple transactions into a single package and moving the computation of these transactions off the blockchain, while still keeping all the transaction data on the blockchain. This can improve security and scalability. One type of Rollup, called zk-Rollup~\cite{zkrollup}, uses zero-knowledge proofs to efficiently compress and verify a large number of transactions. zkSync~\cite{zksync} and Loopring~\cite{Loopring} are platforms that use zk-Rollup. Another type of Rollup, called Optimistic Rollup~\cite{oprollup}, assumes that transactions are legitimate unless proven otherwise through the submission of fraud proofs. Optimism~\cite{Optimism} and Arbitrum~\cite{Arbitrum} are platforms that employ Optimistic Rollup.

\subsubsection{Cross-Chain Techniques} \label{sec:Cross-Chain Techniques}
 Current cross-chain schemes can be divided into chain-based ones and bridge-based ones. 

Chain-based cross-chain schemes refer to those that utilize the chain's own mechanism without requiring other entities such as sidechains~\cite{SINGH2020102471} and hashed timelock contracts (HTLC)~\cite{poon2016bitcoin}. 
A sidechain is a blockchain that is connected to a main blockchain, also known as the master chain. It is considered as a slave chain because it depends on the main chain for its security and functionality. 
Pegged Sidechain~\cite{PeggedSidechain} is a technique that enables cross-chain transactions using a two-way peg method. In a symmetrical setting, assets on the main chain can be transferred to a special output that locks the assets and generates a Simplified Payment Verification (SPV) proof for a particular sidechain. 
Proof-of-Stake Sidechains~\cite{PoSSidechain} and Proof-of-Work Sidechains~\cite{PoWSidechain} are two alternatives proposed by the same research group for handling cross-chain transactions. They both generalize the definition of sidechains so that cross-chain transfers can be performed between any two chains, not just between parent and child chains. In Proof-of-Work Sidechains, the cross-chain proof required for the cross-transfer is generated using Non-Interactive Proofs of Proof-of-Work (NIPoPoWs), while in Proof-of-Stake Sidechains it is generated using Ad-hoc Threshold Multi-signatures (ATMS). 

As a type of chain-based cross-chain scheme,  HTLC is originated from Bitcoin's lightning network and is based on a smart contract with a hash lock and a time lock. The purpose of the hash lock is to lock the corresponding assets in a contract, while the time lock sets a constraint on the locked assets, specifying that the assets can only be withdrawn within a certain time limit, to ensure the atomicity of the scheme. The goal of HTLC is to exchange different blockchain assets between two users across platforms, e.g., exchanging Ether for Bitcoin. 
MAD-HTLC~\cite{tsabary2021mad} effectively resists incentive manipulation attacks through its blockchain-based incentive mechanism. 
XCLAIM~\cite{zamyatin2019xclaim} reduces HTLC's strong assumptions such as the requirements for both parties to be online and for clocks to be synchronized. 
Cross-Channel~\cite{crosschannel} is the first off-chain channel that supports cross-chain services using a hierarchical structure with a settlement protocol and a fair exchange protocol. 
Thyagarajan \textit{et al.}~\cite{thyagarajan2022universal} proposed an approach to use VTS (Verifiable Timed Signature) instead of time locks, which makes HTLC no longer dependent on smart contracts, thereby improving universality. 

Bridge-based cross-chain schemes require the introduction of a bridge to help the chains interact. Example bridge schemes include notary~\cite{Interledger}, relay chains, and DPKC (distributed private key control)~\cite{belchior2021survey}.
A notary scheme, which is popular in the Interledger project, involves finding a trusted third party that both sides of a cross-chain interaction trust, to verify and forward transactions. The implementation of a notary scheme is simple, but it has a single point of failure problem. 
Yin \textit{et al.}~\cite{yin2022bool} developed an open, distributed notary cross-chain platform named Bool Network, which combines security hardware (e.g., Intel SGX) and cryptographic technologies (e.g., MPC and NIZK) to ensure security. 
The relay chain aims to construct a third-party public chain that connects other chains in a blockchain network through a cross-chain message-passing protocol. It essentially redirects transactions from one chain to another through a trusted relay. Cosmos~\cite{Cosmos} and Polkadot~\cite{wood2016polkadot} are representative platforms based on the relay chain mechanism. Compared to each other, Cosmos has an advantage in scalability, but Polkadot is more secure. A more detailed introduction to both will be presented in Section~\ref{Sec:Hieracical}. As a scheme based on Cosmos, zkBridge~\cite{ZkBridge} was proposed in 2022, which employs zero-knowledge proof technology to enhance its system security. 
DPKC is a cross-chain technology based on secure multi-party computation and threshold keys. It uses nodes in a distributed network to control the private keys of accounts storing digital assets on a blockchain, separates the use rights and ownership of digital assets, and maps the original assets on one chain to another chain, thereby enabling asset exchanges between different blockchain systems. Current example schemes based on DPKC include Fusion~\cite{fusion} and Wanchain~\cite{wanchain}.

\subsection{Modular Blockchain Analytic Framework}
\label{sec:framework}

From a macroscopic standpoint, it is evident that the blockchain research community has invested a considerable effort into scaling blockchain systems and maintaining their security properties. Currently, the prevailing trend in blockchain development involves integrating it with the core layers of a digital infrastructure to provide support for an even wider array of applications, such as NFT, Web3.0, and Metaverse. However, to fully comprehend the intricacies of blockchain, it is necessary to analyze it from a microscopic and systematic perspective. For this purpose we propose the modular blockchain analytic framework in this section, which can facilitate the creation of scalable, flexible, and application-specific blockchain systems. Then we use bitcoin as an example to illustrate how this framework can be employed to dissect a blockchain system. 

\subsubsection{Overview}
\begin{figure}[!htbp]
\centering
\centerline{\includegraphics[width=0.7\textwidth]{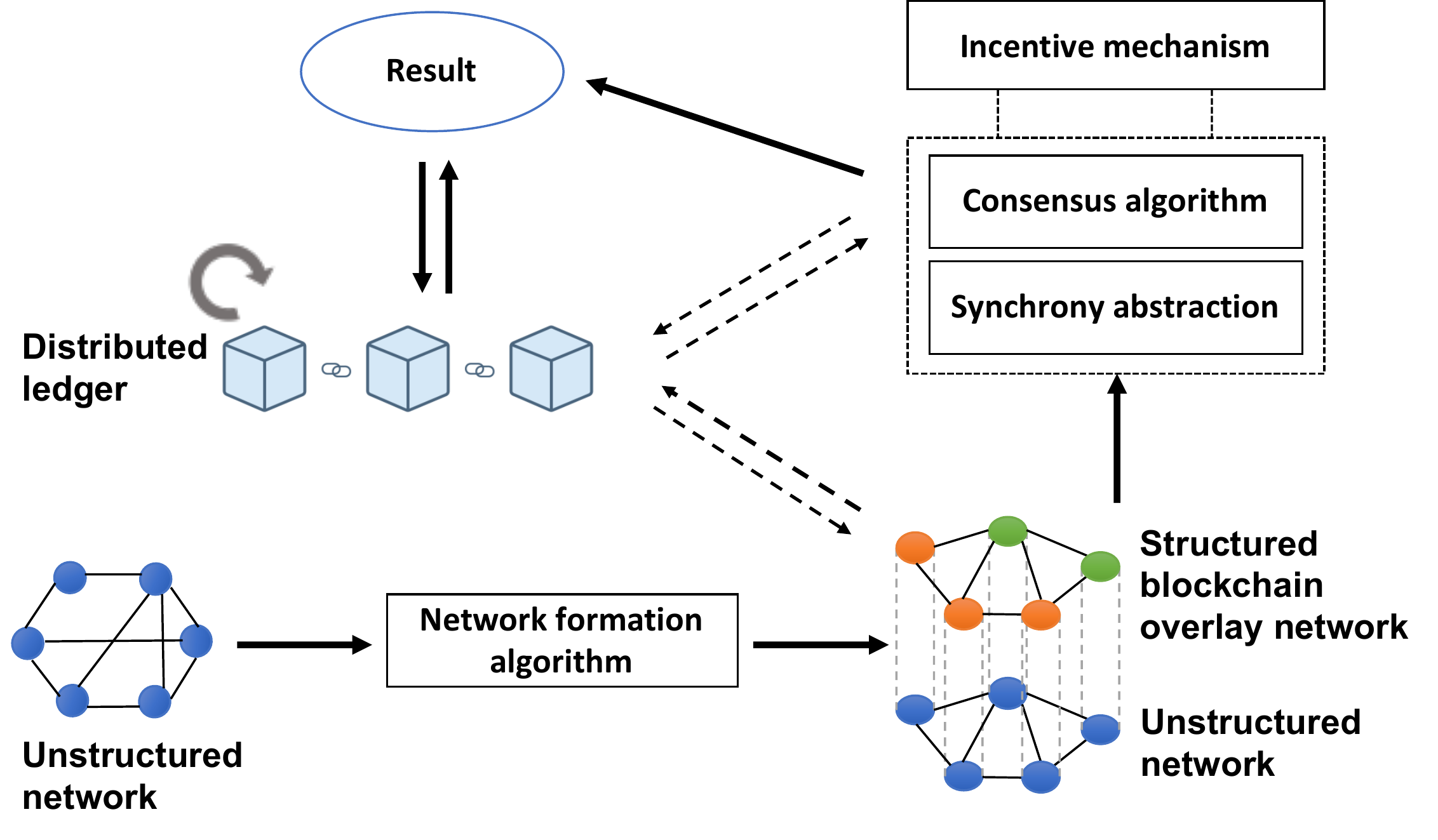}}
\caption{Modular Blockchain Analytic Framework}
\label{Analysis:Model}
\end{figure}

The modular blockchain analytic framework is a general one created for analyzing existing blockchain systems and guiding the design of new ones. This framework decouples a blockchain system to facilitate in-depth analysis. As shown in Fig.~\ref{Analysis:Model}, the network formation algorithm takes an unstructured network as input and produces a structured blockchain overlay network consisting of blockchain nodes. These nodes run a consensus protocol to agree on who should produce the next block and verify the legitimacy of the new block. The new block is temporarily appended in the distributed ledger after validation, waiting to be confirmed later. The incentive mechanism rewards or penalizes blockchain nodes for maintaining the stability and sustainability of the whole blockchain system.

\subsubsection{Proof of Concept}
\label{proof:of:concept}
We use the modular blockchain analytic framework presented above to analyze Bitcoin as an example to provide readers with a guided tour. 

Bitcoin is a decentralized payment system that was introduced in 2008 by Satoshi Nakamoto and was launched in early 2009. It operates without a central bank and allows users to freely join in and leave. Bitcoin's structured blockchain overlay network is formed by a registration process, during which a node can become a miner. The overlay network is therefore ``flat'', with miners serving as consensus nodes and having identical functionality. Users can make payments by digitally signing transactions, which are further ordered by miners (satisfying the total order property) and recorded on Bitcoin's blockchain. Before broadcasting a newly 
formed block, a miner must solve a hash puzzle to provide proof of work, namely Proof-of-Work (PoW) consensus (a.k.a. Nakamoto consensus). PoW is close to Hashcash, relying on a moderately hard computation~\cite{dwork1992pricing}. The hash puzzle in Bitcoin requires that the miners find a nonce such that the concatenation of the nonce, the previous hash, the timestamp, and other data in the block header, hashed twice using SHA-256, is smaller than a target value. The idea behind Proof-of-Work is to randomly select a leader with low probability on average, based on computational power, to decide the next block. A block, as a result of the PoW consensus, is appended to the chain of blocks, which is stored on all nodes. Typically, it takes about one hour to confirm a block in Bitcoin. This design preserves the persistence and liveness properties and thus protects bitcoin from double-spending attacks and Sybil attacks. 
Bitcoin also has two types of incentive mechanisms: the new block reward and the transaction fee, with the former being offered to miners who create a new valid block, and the latter being included in each transaction to incentivize miners to include the transactions in the next block. 

In a nutshell, a node joins the flat Bitcoin network via registration to become either a full node (participating in the consensus process and storing the full blockchain) or a light one (only storing the chain of block headers). A miner, as a consensus node, runs the PoW consensus algorithm to compete for the right of constructing the next block. A new block is appended to the current blockchain after being verified by all miners and later confirmed into the chain. Bitcoin adopts two incentive mechanisms to promote consensus. The distributed ledger of Bitcoin is precisely a chain of blocks.

In the following sections, we delve into each of the framework's key components.

\section{Blockchain Network}
\label{architecture}
\begin{figure}[!htbp]
\centering
\centerline{\includegraphics[width=0.7\textwidth]{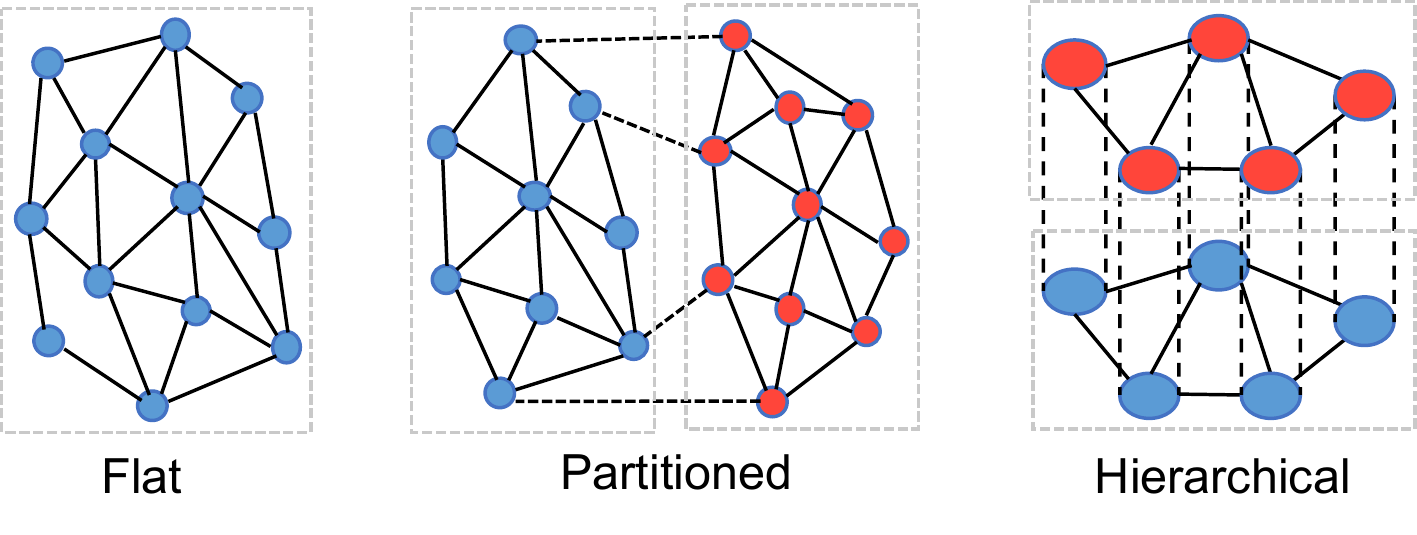}}
\caption{Flat, Partitioned, and Hierarchical networks. Nodes in each gray dashed frame are connected and can interact with each other (e.g., in the same P2P network).}
\label{map:architecture}
\end{figure}

Before the consensus process and ledger update take place, a blockchain network must be established. Generally, the topology of a blockchain network can be \textit{Flat}, \textit{Partitioned} or \textit{Hierarchical}, as shown in Fig.~\ref{map:architecture}. 

\subsection{Flat and Partitioned} \label{sec:Flat and Partitioned}
A flat blockchain network consists of all consensus nodes (also known as validators, miners, etc., based on the context) with identical functionality that forms a connected graph. This topology is prevalent in early blockchain systems such as Bitcoin, Ethereum, Namecoin, Litecoin, and IOTA Tangle. The creation of a flat network is typically achieved through a straightforward registration process that selects a group of consensus nodes. For instance, in Bitcoin, an individual must prepare mining hardware or secure cloud mining contract, set up a wallet, download the mining software, and select a mining pool to join to become an eligible miner. Similar registration processes exist for Ethereum and other altcoins that have a flat network architecture.

A partitioned network implies that consensus nodes are not well-connected. Partitioning can be categorized into two primary types: horizontal and vertical. Horizontal partitioning entails dividing the network into several sub-networks that operate in parallel and asynchronously, yet can still reach a universal agreement. Network partitioning also appears when some nodes disconnect from the blockchain for a while making the network having an inconsistent view during that time. On the other hand, vertical partitioning, which is also known as cross-chain, involves multiple blockchain projects that need to collaborate. In general, horizontal partitioning is adopted by sharding protocols while vertical partitioning is used in cross-chain systems.

Ripple, ELASTICO, Omniledger, and Monoxide are some examples that take horizontally partitioned network topologies. Ripple divides its blockchain nodes into fully connected shards called cliques and uses RPCA to ensure that any two shards have at least a certain number of nodes overlapping in order to prevent double-spending attacks. ELASTICO addresses scalability and communication issues using the PBFT consensus algorithm within different shards. Omniledger partitions validators into different shards with a focus on improving the security of the partitioning process by making it unpredictable, bias-resistant, and verifiable. Monoxide addresses the issue of lock/unlock overhead by introducing eventual atomicity and addresses the dilution problem of mining power or stake share in sharding with a technique called Chu-ko-nu mining. This technique distributes the mining power among multiple shards by requiring miners to mine in different zones simultaneously but no more than one block per zone can be produced, thereby increasing the security of each shard and revitalizing the PoW consensus algorithm in a sharded blockchain.

In vertical partitioning, each blockchain independently maintains its own network. When supporting cross-chain interoperations, the entire network can be viewed as being vertically divided into multiple sub-chain networks.
To be specific, each chain does not need to store the complete ledgers of other blockchains, and the interacting parties achieve atomic interactions through a cross-chain protocol. Some cross-chain protocols assist in completing the interaction by introducing trusted entities on both sides, such as notaries in the notary scheme. These schemes are often easier to implement but introduce centralization. Distributed Private Key Control (DPKC) and relay chains to some extent alleviate the single point of failure problem but are difficult and costly to develop. The distributed network of DPKC is stable and secure, while the nodes in the relay chain can be dynamically fault-tolerant. Other schemes, such as HTLC, aim to guarantee the atomicity of the protocol without introducing extra entities. Recently, multiple schemes have emerged to improve HTLC and aim to achieve a general distributed cross-chain protocol (more details are presented in Section~\ref{sec:Cross-Chain Techniques}). However, these techniques are currently still in the experimental phase and their feasibility for implementation remains to be tested. 

\subsection{Hieracical} \label{Sec:Hieracical}
There are two main categories of hierarchical blockchain networks: committee-based and multi-blockchain. A committee-based network selects at least one committee based on certain criteria such as identity, stakeholding, or randomness. This committee usually shepherds other consensus nodes to reach a consensus, forming a hierarchy. Multi-blockchain networks, on the other hand, contain multiple blockchains that work together in a hierarchical manner, with at least one beacon blockchain coordinating the others. These blockchains can be regarded as forming a tree-like hierarchical structure. It is important to note that partitioned and hierarchical networks are not mutually exclusive but sometimes complement each other. In other words, a blockchain network can have both partitioned and hierarchical design elements.

We use Algorand, ELASTICO, and Byzcoin as illustrations to clarify the concept of committee-based networks.
The network of Algorand is hierarchical due to the existence of a committee. Algorand adopts the Verifiable Random Function (VRF)~\cite{micali1999verifiable} to randomly select a committee. The VRF can guarantee that the committee is selected confidentially. 
ELASTICO combines partitioned and hierarchical designs. It establishes a final committee made up of individuals who are randomly selected from each shard. The final committee is responsible for confirming new blocks submitted by shards.
Byzcoin presents a dynamic committee system that can utilize both PoW and PBFT, and introduces a PBFT-based consensus mechanism for strong consistency. 

Several cross-chain systems (adopting a vertical partitioning architecture shown in Section~\ref{sec:Flat and Partitioned}) such as Cosmos, Polkadot, and HyperService~\cite{liu2019hyperservice} make use of a multi-blockchain network topology. Cosmos employs a network of independent blockchains called zones, which are connected by a central Cosmos Hub responsible for maintaining the network and enabling horizontal sharding through center-supervised interoperability. The hub and zones communicate with each other using an inter-blockchain communication (IBC) protocol based on secure atomic token exchanges via the hub. Polkadot uses a relay chain to support heterogeneous multi-chains called parachains, with the relay chain serving as a beacon chain to bridge the parachains. The relay chain's consensus is driven by the Parity Substrate and operated by a validator swarm consisting of validators designated to different parachains. For cross-chain communications, validators on the relay chain use a queuing mechanism for routing transactions and store the transactions on the parachains after referencing them on the relay chain. HyperService is a platform for interoperability and programmability across heterogeneous blockchains, featuring a Network Status Blockchain (NSB) that provides an objective and unified view of the status of all blockchains. The NSB's blocks contain StatusRoot and ActionRoot, the roots of two extra merkle trees, which respectively record transaction status and actions of the underlying blockchains.

\section{Consensus Protocol}
\label{consensus}

Consensus is a critical aspect of blockchain operation in which all nodes participate, agree on the same result, and output a unanimous global view. A consensus protocol can be broken down into three components: synchrony abstraction, consensus algorithm, and incentive mechanism. Although these components are interdependent, separating them is beneficial since it enables us to examine the consensus protocol from three distinct perspectives: setting the scene, achieving consensus, and ensuring sustainability. Synchrony abstraction deals with the underlying assumptions of the consensus protocol, the consensus algorithm determines how to attain consensus, and the incentive mechanism concentrates on maintaining the sustainability of the process.

\subsection{Synchrony Abstraction}
\label{synchronous}
Before analyzing a blockchain consensus algorithm, we first figure out its synchrony abstraction (also referred to as synchrony assumption, or timing assumption). Generally, a blockchain network can be synchronous, partially synchronous, or asynchronous, which are detailed as follows. 

\begin{itemize}
\item \textit{Synchronous (Fully-synchronous and Semi-synchronous):} Under the synchronous setting, processes execute in lock-step, and message transmission delay is $\Delta$-bounded, where $\Delta\in(0,1]$ is a known fixed upper bound. Some blockchain systems even make a strong assumption that network channels are \textit{fully-synchronous} without delays, i.e., $\Delta=1$. In a \textit{semi-synchronous} model, $\Delta\in(0,1)$, which can be achieved using a universal clock or relying on a good network connection. The synchronous setting is convenient for theoretical analysis in that synchrony leads to deterministic time and order of blocks, but real-world cases are at best partially synchronous and generally asynchronous.
\item \textit{Partially synchronous:} The partially synchronous model lies between synchronous and asynchronous models~\cite{dwork1988consensus}. There are two situations under which one can claim that a blockchain is partially synchronous. First, an upper bound of the message transmission delay, i.e., $\Delta$, exists but unknown; here $\Delta$ is also known as the \textit{a-prior} bound. Second, the $\Delta$ exists and is known, but the blockchain system is unreliable, and the $\Delta$ only partially takes effect.
\item \textit{Asynchronous:} Message transmission delays are unbounded. In this circumstance, consensus can not be reached in the fail-stop case that terminates within a bounded time, even with only one crash failure according to the FLP impossibility result (named after Fischer, Lynch, and Patterson)~\cite{fischer1982impossibility}. Thus consistency can not be satisfied either. 
\end{itemize}

Researchers have studied the essential properties of public ledgers under both synchronous~\cite{garay2015bitcoin} and partially synchronous settings~\cite{pass2017analysis} to ensure the proper functioning of a blockchain. These properties include persistence and liveness, which can be further categorized into chain growth, chain quality, and common prefix. Different types of consensus algorithms consider various properties with slight differences. For instance, protocols based on Byzantine Fault Tolerance - State Machine Replication (BFT-SMR) aim to ensure agreement, system functionality, and total order of events, while protocols based on Byzantine atomic broadcast (BAB) guarantee agreement, integrity, validity, and total order. Both BFT-SMR and BAB can serve as building blocks for blockchain systems.

\subsection{Proof-of-X Consensus}\label{consensus:pox}
\begin{figure}[htbp]
\centering
\centerline{\includegraphics[width=0.7\textwidth]{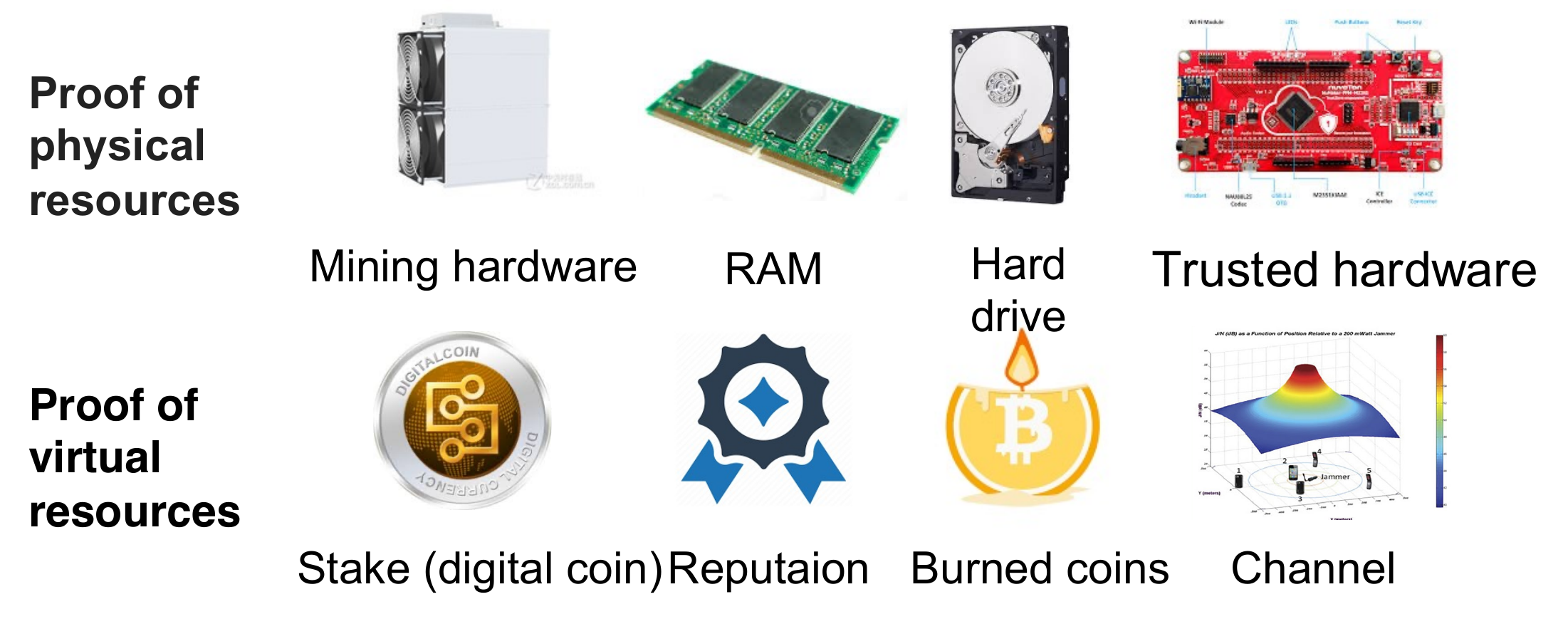}}
\caption{Proof-of-X consensus.}
\label{fig:pox}
\end{figure}

The Proof-of-X consensus is a set of protocols used in blockchain to determine which consensus node should act as the leader to propose a block. Nodes demonstrate their qualifications by utilizing their resources, and the selected node becomes the leader responsible for validating transactions and adding them to a new block, which is then shared with others. Fig.~\ref{fig:pox} illustrates two categories of PoX consensus protocols: those that rely on physical resources, such as mining hardware, RAM, hard drive, or trusted hardware, and those that rely on virtual resources, such as stake, reputation, burned coins, or communication channels. Nodes compete with each other to prove their uses of these resources to gain the opportunity of proposing a block.

In PoW-based consensus algorithms, miners compete with each other to add new blocks to the blockchain using their computational power. A miner with more computational power has a higher probability of proposing the next block. Bitcoin is the first global decentralized transaction ledger that employs a PoW consensus algorithm based on the hashcash with double-iterated SHA-256. Ethereum makes use of a PoW algorithm called Ethash, which is rooted in the Dagger-Hashimoto algorithm and designed to be memory-hard. This means that finding a nonce requires a lot of memory and bandwidth, making it difficult for parallel computing. To address the scalability issues in Bitcoin, Bitcoin-NG separates Bitcoin's PoW-based consensus process into two sub-processes: leader election and transaction serialization, with the former still utilizing a PoW-based consensus algorithm to agree on a key block, whose proposer is responsible for the next epoch.

Other proof of physical resource mechanisms include Proof-of-Space (PoSpace)~\cite{pospace}, Proof-of-Elapsed-Time (PoET)~\cite{chen2017security}, and Proof-of-Retrievability~\cite{bowers2009proofs}. PoSpace, also known as Proof-of-Capacity or Proof-of-Storage, involves nodes competing based on the memory or disk space they use. This concept can be applied to design consensus algorithms for nodes with sufficient storage hardware. PoSpace is also prevalent in decentralized storage networks, e.g., Filecoin, Storj, Sia. PoET requires each node to generate a random number to determine how long it must wait before being allowed to generate a block. It was proposed by Intel and uses trusted hardware, such as SGX, to enforce the random waiting time. Proof-of-Retrievability in the context of the file system refers to a method used by a file system (prover) to prove to a client (verifier) that a specific file is complete and can be fully recovered. This concept can be used to create consensus algorithms for nodes that have storage systems, such as cloud servers. For example, Permacoin~\cite{miller2014permacoin} is a consensus algorithm that requires nodes to invest both storage and computational power to compete to write blocks. 

The inefficiency of the PoW consensus mechanism due to its energy consumption has prompted the development of the proof of virtual resource concept. Despite being a widely adopted consensus mechanism, PoW wastes resources as many miners work on hard computational problems for the same block simultaneously, resulting in high computing power and energy usage. In contrast, proof-of-stake (PoS) was introduced as an alternative in 2011. In PoS, the creator of a new block is selected based on the stakes the miners hold in the network. The term ``stake'' refers to the number of coins held by a node and committed to the network as collateral, and a higher stake results in a higher probability of being selected as the next block creator. Other examples of proof of virtual resources include the Delegated Proof of Stake (DPoS)~\cite{Dpos}, Proof-of-Activity~\cite{Activity}, Proof-of-Authority (PoA)~\cite{Authority}, Proof-of-Reputation (PoR)~\cite{gai2018proof}, Proof-of-Burn (PoB)~\cite{proofofburn}, and Proof-of-Channel (PoC)~\cite{BLOWN, wChain, zou2021fast}. 

Delegated Proof of Stake (DPoS) is a variant of PoS in which nodes (called ``delegates'') are elected by the token holders to validate transactions and create new blocks. The elected delegates are also responsible for maintaining the network, and get rewarded for their work with a portion of the transaction fees and block rewards. DPoS was designed to be more efficient and scalable than the traditional PoS, as the number of delegates is typically much smaller than that of the nodes in the network. 
Proof-of-Activity combines PoW and PoS, allowing a blockchain system to switch between the two to mitigate the problems of either PoW or PoS. 
Proposed by Gavin Wood, PoA (Proof of Authority) is a consensus mechanism that allows only approved authorities or validators to propose blocks, and it was specifically designed for permissioned blockchain systems.
PoR is a variation of PoA, where each node is assigned a reputation score and only the nodes with a reputation above a certain threshold are allowed to create blocks. Compared to PoA, PoR is considered more secure and dependable.
PoB was proposed by Iain Stewart in 2012. It relies on destroying coins to virtually earn the right to write new blocks. Slimcoin~\cite{slimcoin} uses PoB as its consensus algorithm. With proof-of-channel (PoC)~\cite{BLOWN}, nodes are required to send messages or perform other types of communication-based tasks to prove their efforts and potentially be chosen as the leader for the next block in wireless networks. 
PoC intends to integrate consensus and channel competition seamlessly. It involves nodes competing for channels to propose blocks, which incorporates channel conditions into the consensus process. This approach reduces communication costs and improves the efficiency and effectiveness of consensus in wireless networks under conditions of adversarial jamming and variant channel bandwidth.

\subsection{BFT Consensus}\label{consensus:bft}

BFT (Byzantine Fault Tolerance) consensus algorithms in blockchain typically pertain to BFT-SMR, e.g., PBFT, Tendermint, and Hotstuff. However, in addition to BFT-SMR, there exists a collection of BFT protocols such as byzantine reliable broadcast, byzantine atomic broadcast, asynchronous byzantine agreement, multi-value byzantine agreement, and asynchronous common subset, which can serve as foundational elements for constructing BFT consensus algorithms. 
The concept of Byzantine Fault Tolerance refers to the reliability of a fault-tolerant computing system, particularly for distributed computing systems~\cite{distler2021byzantine}. The goal of BFT consensus algorithms is to ensure that a blockchain can withstand system failures and function correctly by reducing the impact of faulty nodes and preserving the consensus reached by the honest majority (at least $n-f$ nodes agree when $n=3f+1$). For example, PBFT, as a BFT-SMR algorithm, has three key phases to achieve tolerance: Pre-prepare, Prepare, and Commit. The Pre-prepare and Prepare phases are used to order requests that are sent in the same view, even when the primary node, which proposes the ordering of requests, is faulty. The Prepare and Commit phases ensure that requests that are committed are ordered across views. A view change protocol is also needed to avoid timeout issues and maintain the liveness of the system.

Many blockchains adopt a BFT consensus algorithm. ByzCoin is a cryptocurrency that uses a consensus protocol called collective signing, termed CoSi. In ByzCoin, key blocks are proposed using a proof-of-work consensus process, while micro blocks are created with the CoSi protocol. The system also includes a dynamic committee that supports both proof-of-work and PBFT (a type of Byzantine fault tolerance protocol) and protects against Sybil attacks. The Ripple network includes a consensus protocol called RPCA, which is run in multiple rounds to finalize a set of transactions. Ripple can tolerate up to 20\% nodes behaving in a Byzantine manner and has a lower security threshold of 80\% of honest nodes. Traditional BFTs like PBFT require a known number of participating nodes and may be vulnerable to Sybil attacks. Tendermint-BFT~\cite{Tendermint} remedies this by adding proof-of-stake on consensus, which only allows validators with enough pledged stakes to participate in the consensus. 
In a partially synchronous network, Hotstuff has a communication complexity that is linear in the number of replicas. This is an improvement over other BFT protocols such as BFT-SMaRt~\cite{SousaB12}, which has a quadratic communication complexity during view change. HotStuff is designed to operate in such a network by allowing a correct leader to drive the protocol to consensus at the pace of actual network delay, which is called responsiveness.

With the continuous developments of the blockchain technologies, a number of new BFT protocols have been proposed for asynchronous networks. These protocols are designed not only for use in blockchains but also for offering other functions such as secret sharing, decentralized trusted setup, and multi-party computation. The recent progress of BFT protocols indicates the trend of the fusion of BFT protocols and cryptographic primitives under decentralized settings. For details, we recommend a comprehensive survey in~\cite{wang2022bft}.

\subsection{Hybrid Consensus} \label{consensus:hybrid}

Hybrid consensus is the combination of multiple consensus algorithms within one protocol to fully utilize their benefits.
Peercoin (PPCoin)~\cite{king2012ppcoin} is the first cryptocurrency to adopt a hybrid consensus algorithm that combines proof-of-work (PoW) and proof-of-stake (PoS). In PeerCoin, the PoS component uses the concept of ``coin age'', which is determined by multiplying the number of coins held by their holding period. The coin age is then employed in the PoW process, with the difficulty of solving the cryptographic puzzle being inversely proportional to the coin age. This allows nodes with larger coin holdings to have a higher likelihood of proposing the next block. Such a hybrid approach offers numerous benefits, such as reducing energy consumption when compared to pure PoW and minimizing the impact of centralized mining pools. It also makes launching a 51\% attack in a PoS system more costly since the attacker would need to control 51\% of all coins. Note that these advantages are not exclusive to Peercoin -- they are applicable to other PoS-based cryptocurrencies such as Nextcoin. Moreover, the hybrid consensus algorithm remains a popular choice for transitioning between different consensus algorithms.

\subsection{Incentive Mechanism}
\label{incentive:mechanism}

An incentive mechanism is a method of motivating individuals or machines to complete tasks accurately and effectively. When it comes to blockchain, an incentive mechanism is closely linked to the consensus protocol. It encourages consensus nodes to act honestly during the consensus process and promotes the security of blockchain. The consensus protocol acts like a car's steering wheel, safely guiding the blockchain system in a specific direction, while the incentive acts as fuel to keep the system operating efficiently and effectively.


Currently, researchers are studying blockchain incentive mechanisms from both negative and positive viewpoints. When looking at it negatively, attackers can take advantage of flaws in an incentive mechanism to launch attacks and reap benefits.
Ittay Eyal and Emin G{\"u}n Sirer~\cite{ittay} identified security issues in the incentive mechanism of proof-of-work-based blockchains and proposed solutions including a new backward-compatible incentive protocol to address them. 
Zur \textit{et al.} proposed PTO in~\cite{zur2020efficient}, which helps miners to optimize their mining strategies and balance the relationship between computing power consumption (in PoW consensus) and rewards.
In~\cite{tsabary2018gap}, the authors analyzed the relationship between mining expenses (related to consensus) and rewards (related to incentives). They further explained that if the accumulated transaction fees do not exceed a certain threshold, miners have no incentive to mine.
In addition, Mirkin \textit{et al.}~\cite{mirkin2020bdos} described a type of incentive-based denial-of-service attack called blockchain denial-of-service (BDoS) attack, which is different from traditional selfish mining attacks as it aims to disrupt the system rather than increase the attacker's revenue.

From a positive viewpoint, developers of blockchain systems can use incentive mechanisms to attract more users and maintain the normal operations of the system. Tedjamulia \textit{et al}. identified three major types of incentives: financial, performance appraisal, and social recognition, which are also applicable in blockchain systems. These incentives can be further divided into financial incentives (e.g., payments and bonuses) and social incentives (e.g., reputation and trust). Financial incentives, such as mining rewards in Bitcoin and Ethereum, are the most commonly used ones in blockchain. The design of these incentives often incorporates techniques from economics, such as game theory~\cite{sun2021applications, liu2019survey}. For example, Asgaonkara \textit{et al}.~\cite{asgaonkar2019solving} proposed a dual-deposit escrow trade protocol for verifiable digital goods that utilizes a game between the buyer and the seller to ensure the safety and liveness of the protocol through the use of sub-game perfect Nash equilibrium and the opportunity cost of locked deposits.

\section{Distributed Ledger}
\label{ledger}

\begin{figure}[htbp]
\centering
\centerline{\includegraphics[width=0.7\textwidth]{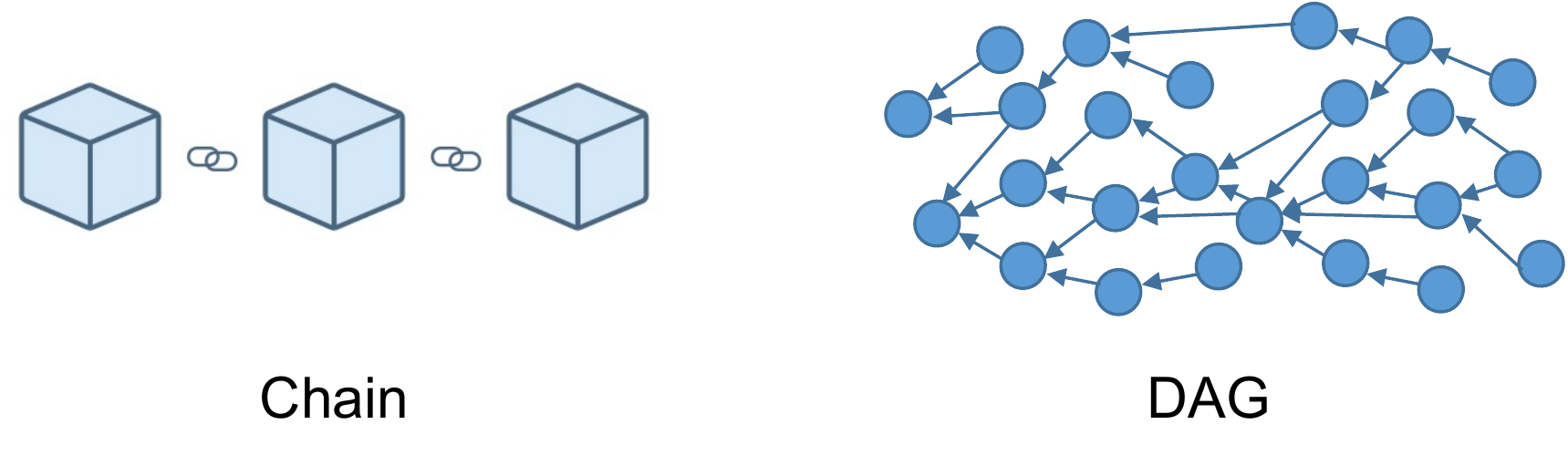}}
\caption{Two types of ledger structures: chain and DAG.}
\label{fig:Ledger}
\end{figure}

There exist two categories of distributed ledgers: chain-based and directed acyclic graph (DAG)-based. How a ledger is stored and maintained is determined by its structure. A chain structure is commonly used for blockchains and has been extensively researched and implemented. DAG-based ledgers, on the other hand, have gained popularity in recent years due to their ability to process transactions in parallel and their low communication/time complexity.

\subsection{Chain}

A chain structure is made up of chained blocks, with each containing a hash pointer that connects it to the previous block. The chain's linear format is useful for organizing and verifying transactions. Such a format is difficult to manipulate because any modification would break the chain, which could be easily detected by checking block hashes. It is important to note that a chain is made up of all confirmed blocks -- it doesn't include pending ones. If pending blocks are considered, the end of the chain may fork temporally. To avoid the fork problem, most blockchain systems have protocols in place to merge forks into a singular chain. Examples of such protocols include the longest chain rule, GHOST \cite{sompolinsky2015secure}, and EOS \cite{EOS}.

The longest chain rule states that when a blockchain is forked, the path with the most blocks and thus the longest chain is considered the correct one. This longest chain reflects the greatest amount of effort (PoW) put in by miners. Assuming the honest majority, it is highly unlikely that an attacker could manipulate the direction of the chain's growth according to Gambler's theory. The longest chain rule ensures that the network safely reaches a consensus on a linear chain without forks and conflicting transactions. However, it suffers from the following issue:
%
as the market share of Bitcoin grows rapidly, it needs to process an increased number of transactions, which brings the problem of limited throughput to light. Nevertheless, it is always challenging to optimize the tradeoff between efficiency and security, due to the fact that the two main methods of increasing throughput - making computational problems easier and increasing block size - would both result in more forks, which could decrease the security strength in protecting against double-spending attacks. 

To address this problem, Sompolinsky and Zohar proposed the GHOST protocol to select the Greedy Heaviest-Observed Sub-Tree instead of choosing the longest chain when dealing with forks \cite{sompolinsky2015secure}. A variant of GHOST has been implemented in Ethereum.  GHOST considers all blocks when calculating the overall PoW, which gives the honest nodes greater influence in the network. With this protocol, attackers cannot easily overtake the network by extending a malicious chain to make it the longest one. With GHOST, a blockchain can obtain higher throughput and become more secure, which cannot be simultaneously achieved under the longest chain rule.

In EOS, a block producer should always expand the longest chain, and must not expand two chains simultaneously. This is done by requiring each transaction to store a hash of a recent block, which means that the producer of this block chooses this chain. Once signed, the producer is forbidden to sign any other chain -- a 
double signing is regarded as a violation. This design is called Transaction as Proof of Stake (TaPoS). It can achieve two goals: (1) preventing replay attacks of transactions on two chains and (2) preventing forking.

\subsection{DAG}

A DAG (directed acyclic graph) is a directed graph without cycles. It has been used to increase the efficiency of chain-based blockchains. Unlike a chain, a DAG can have multiple predecessors for each vertex. Transactions that have been validated can be added directly to a DAG, thereby bypassing the block creation. As a result, some projects that use DAGs do not even consider themselves blockchains, but rather general (or blockless) distributed ledgers. In a DAG, transactions can be appended to the ledger in parallel, leading to high throughput. However, DAGs have three inherent problems that need to be addressed: (1) it is difficult to maintain a universal, consistent view among all nodes in a DAG since achieving the total order property is intriguing,  making it challenging to prevent double-spending and forks in the DAG; (2) designing and building a secure DAG is challenging due to its complex two-dimensional structure, rendering it hard to conduct security analysis and deploy DAG-based blockchains in practice; and (3) if there are conflicts between two transactions, it can take a large amount of time to verify them in a DAG, which may cause significant delays in transaction processing.

Some early examples of DAG systems include the IOTA Tangle, Byteball, and Fantom. The IOTA Tangle is the first widely accepted DAG-based ledger. In Tangle, when a transaction is issued, its owner must validate two previous transactions. If there exists no conflict with the Tangle's history, the new transaction is attached to the two selected previous ones as approval. The new transaction then waits for the next new transaction's approval, which may result in a long finalization time. Transactions in Tangle do not form batches or blocks -- they are the smallest unit. To prevent spam and Sybil attacks, Tangle uses a proof-of-work nonce similar to that in Bitcoin to make it more secure. As discussed in Section~\ref{dag}, there also exist a series of blockDAG systems, including the Inclusive Block Chain Protocols, SPECTRE, and PHANTOM GHOSTDAG, in which vertices represent blocks instead of transactions. 

Early DAG-based designs usually do not have a formal analysis on their consensus protocols, especially under an asynchronous network setting. Transactions in a DAG do not have a deterministic time upper bound for reliable finalization\footnote{Some Inclusive Block Chain Protocols call this DAG's forgiving nature.}, which makes it difficult to ensure the security properties such as double-spending prevention, transaction atomicity, and unforgeability, in a blockchain. 
It also remains unknown whether some DAG-based blockchains satisfy the total order property. A classic chain-based blockchain maintains the total order of all transactions, which allows for deterministic verification of new transactions. However, DAG structures with only a partial order of their transactions might have vulnerabilities that verification cannot be done deterministically, making it hard to guarantee the prevention of double-spending attacks. 

Some methods of achieving total order have been proposed. Byzcoin \cite{ByzCoin} and Bitcoin-NG \cite{Bitcoin-NG} attempt to establish a main chain hooking sub-chains in order to achieve a full order, but this design does not fully benefit from the parallel processing since the confirmation latency of the blocks on sub-chains is limited by the growing speed of the main chain. 
In Conflux \cite{conflux}, the Tree-Graph consensus mechanism includes a parent selection process which is used to determine the total order of the transactions in the blockchain. When a node receives a new transaction, it first selects one of its relatives as the parent of the transaction and then attaches the transaction to the parent. This Tree-Graph structure ensures the order of transactions across the network and each node can rebuild the ledger by following this structure, thereby the full order of the ledger is maintained. 
The DAG-Rider protocol \cite{dag-rider} divides a DAG into waves, with each containing four consecutive rounds. A leader is randomly selected in the first round and confirms the vertices (each vertex in the DAG-Rider represents a transaction) at the end of the wave. The DAG determines the order of vertices based on their connectivity. Each wave actually contributes to one ``block'', which serializes transactions from four rounds.

\section{Other Techniques and Popular Applications}
\label{sec:techniques}

\subsection{Other Techniques Integrated with Blockchain}

\subsubsection{Smart Contracts}
\label{smart:contract}

In 1994, Nick Szabo proposed the concept of smart contracts as computerized protocols that execute complex term structures in a standardized contract. This idea was revisited with the emergence of cryptocurrencies. Bitcoin allows for the use of a script, which is a stack-based language that extends the capabilities of the Bitcoin protocol~\cite{andrychowicz2014fair,bentov2014use,kumaresan2015use}. However, the Bitcoin script has limitations in its ability to support sophisticated protocols due to its lack of Turning completeness and fine-grained manipulations. Specifically, scripts are inefficient in computing loops and not user-friendly for programmers. Additionally, a script only supports a coarse-grained manipulation of the variables, so fine-grained and multi-stage state transitions are not possible~\cite{buterin2014next}. To address these issues, blockchain systems such as Ethereum, Ripple, Hyperledger, Arbitrum, and Enigma, come with smart contracts that use turning-complete languages\footnote{Smart contracts are Turing complete in theory, but not Turing complete in practice due to the gas limit.} and virtual machines.

Ethereum, which was launched in 2015, has an exciting module called the Ethereum Virtual Machine (EVM), which is a turning-complete machine that provides a runtime environment for smart contracts. A piece of EVM code is a set of bytecode instructions that can be run on an EVM to change the state of the Ethereum network (regarded as a large state machine). The cost of any computation on the EVM is universally determined by pre-defined formulas in units of \textit{gas}. For example, adding two 64-bit integers costs 100 gas. \textit{Ether} is the currency used in the Ethereum network to buy gas and pay for computation efforts. A state in Ethereum is made up of accounts that contain a nonce, ether balance, contract code, and storage. Accounts can be either externally owned or contract accounts. Externally owned accounts do not have code and are used to show balance and issue transactions, while contract accounts have code and control local storage. This allows Ethereum to extend the capabilities of distributed payment systems to a distributed computing platform, but it also introduces new problems and challenges. The EVM and the sophisticated smart contracts lead to new security issues, as exemplified by the DAO hack, in which 97\% of ETH holders reached a consensus to hard fork and undo the theft of about 1.5 million ether. Additionally, Ethereum's proof-of-work (PoW)-based consensus algorithm and fee schedule policy make it expensive to run large computations due to the high gas cost. As a result, applications based on Ethereum are limited by their performance. To address these issues, Ethereum is switching to a proof-of-stake (PoS)-based consensus algorithm. 

Arbitrum makes use of a virtual machine (VM) to implement smart contracts and an incentive mechanism to reach an agreement off-chain regarding what the VM should do. This addresses the miners' free-riding issue due to the high cost of computation for verification and the possibility of a greedy miner consuming others' execution time for profit by including a transaction with a heavy workload. An alternative solution is to implement a participation game like TrueBit, but this can lead to the risk of a participant launching a Sybil attack by registering multiple verifier identities. 
%
The Enigma protocol~\cite{Enigma} offers secret smart contracts programmed in Rust and running on a modified Ethereum Virtual Machine. These smart contracts protect information from the public by encrypting inputs and outputs. Only the application user can execute the contract, and the workload is assigned to a worker node that computes using a modified Web Assembly interpreter (WASMI) running in a Trusted Execution Environment (TEE). The TEE protects the computation process and provides a trusted cryptographic proof as an evidence of a successful completion and verification of the task and its results.

There are ways to improve the efficiency of smart contracts. One of these methods is implemented by Saber, which employs parallel and asynchronous executions to scale smart contracts, as discussed in the article by Liu \textit{et al.}~\cite{liu2021parallel}. By separating the consensus and execution processes, Saber can delegate resource-intensive execution tasks to multiple execution nodes that can work in parallel. Additionally, the nodes can collaborate asynchronously to execute complicated transactions in a non-blocking manner. 
The latency-first smart contract is a method that optimistically accepts committed transactions during high-demand periods, allowing users to complete the verification of the transactions when the blockchain is in low demand~\cite{qilatency}. This approach reduces latency by temporarily overclocking the system, improving efficiency, and freeing up resources for other transactions.

\subsubsection{Zero-knowledge Proofs} \label{sec:Zero-knowledge Proofs and MPC}

Zero-knowledge proof (ZKP) is a protocol through which a prover can convince a verifier that it knows a value without revealing any related information. ZKP can be beneficial for blockchain systems in two ways: hiding sensitive information to protect user privacy, such as identity and privacy-preserving transactions, and simplifying the verification process to improve efficiency, for example, by using a ZKP-based authentication. 

ZKP has been implemented in several popular blockchain systems. Zerocoin~\cite{miers2013zerocoin} and Zerocash~\cite{sasson2014zerocash} (the corresponding implementation is Zcash~\cite{hopwood2016zcash}) employ ZKP to provide anonymity guarantees by preventing transaction graph analysis. Zerocash allows users to send anonymous transactions with hidden amounts and reduces the size and verification time of transactions. Quorum, a blockchain platform developed by J.P. Morgan, supports private smart contracts with hidden business logic and uses ZKP to address double-spending attacks through the Zero-knowledge Security Layer (ZSL). Hyperledger Fabric uses ZKP for anonymous authentication through its Identity Mixer and for privacy-preserving asset exchanges through Zero-Knowledge Asset Transfer (ZKAT).
Monero~\cite{Monero} is a privacy-focused blockchain system that uses a UTXO model and employs ring signatures~\cite{rivest2001leak} and Bulletproofs~\cite{bunz2018bulletproofs} to protect the privacy of both participants and transaction amounts. Solidus~\cite{cecchetti2017solidus} is a blockchain system that follows the account model and is suitable for confidential transactions on public blockchains. It uses Generalized Schnorr Proofs to achieve public verification of on-chain transactions. zkLedger~\cite{narula2018zkledger} is a blockchain system specifically designed for banks, which supports private transactions and public auditing using Schnorr-type non-interactive zero-knowledge proofs. BlockMaze~\cite{guan2020blockmaze} is a blockchain system based on the account model rather than the UTXO model, and uses the zero-knowledge balance and zk-SNARKs to hide transaction amounts and relationships. zkRollup is a layer 2 scaling solution based on zero-knowledge proofs, which performs complex calculations and proof generations off-chain while verifing proofs and storing partial data on-chain to ensure data availability.

\subsubsection{Trusted Hardware}

Trusted hardware components have been utilized in blockchains to enhance security and facilitate new designs. The meaning of trusted components is broad and contentious. This article mainly focuses on two major trusted hardware solutions: Trusted Platform Module (TPM) and Trusted Execution Environment (TEE). TPM is a hardware component with a rigid definition and exclusive implementation by the ISO and the Trusted Computing Group (TCG). It is physically isolated from the rest of a system and performs specific cryptographic computations without allowing any internal programming or modification. TEE (e.g., Intel SGX, Arm TrustZone) is an area on a chipset that provides secure computation services similar to TPM, but developers can program the exact implementation.

The following studies have employed TPM and TEE.
Hardjono \textit{et al.}~\cite{hardjono2016cloud} made use of Enhanced Privacy ID (EPID) TPM to enable anonymous identities in permissioned blockchains. Smith \textit{et al.}~\cite{smith2019methods} employed a TPM secure boot module to build a secure blockchain client that can verify and add external ledger transactions. Jesus \textit{et al.}~\cite{jesus2018blockchain} used a virtual TPM to create a root-of-trust in a blockchain system. 
Milutinovic \textit{et al.}~\cite{milutinovic2016proof} took Intel SGX to create a secure Proof of Luck Consensus Protocol as an alternative to Proof-of-Work. Lind \textit{et al.}~\cite{lind2017teechain} made use of Intel SGX to develop an off-chain payment protocol for efficient, secure, and scalable fund transfers on top of a blockchain. 
Liu \textit{et al.}~\cite{TEMS} introduced a framework that extends trust from on-chain to off-chain. This framework involves a system that uses sensors connected to TEE to continuously monitor the environment and generate anti-forgery data. It also includes a consistency protocol that allows the environment status data to be uploaded from the TEE system to the blockchain in a way that is truthful, real-time, and fault-tolerant. Ayoade \textit{et al.}~\cite{ayoade2018decentralized} employed TEE to verify the integrity of local data storage for decentralized IoT data management.

\subsection{Applications}
\label{sec:app}

\subsubsection{Cryptocurrencies and NFT}

\begin{figure*}[htbp]
    \subfigure[]{\includegraphics[width=0.49\textwidth]{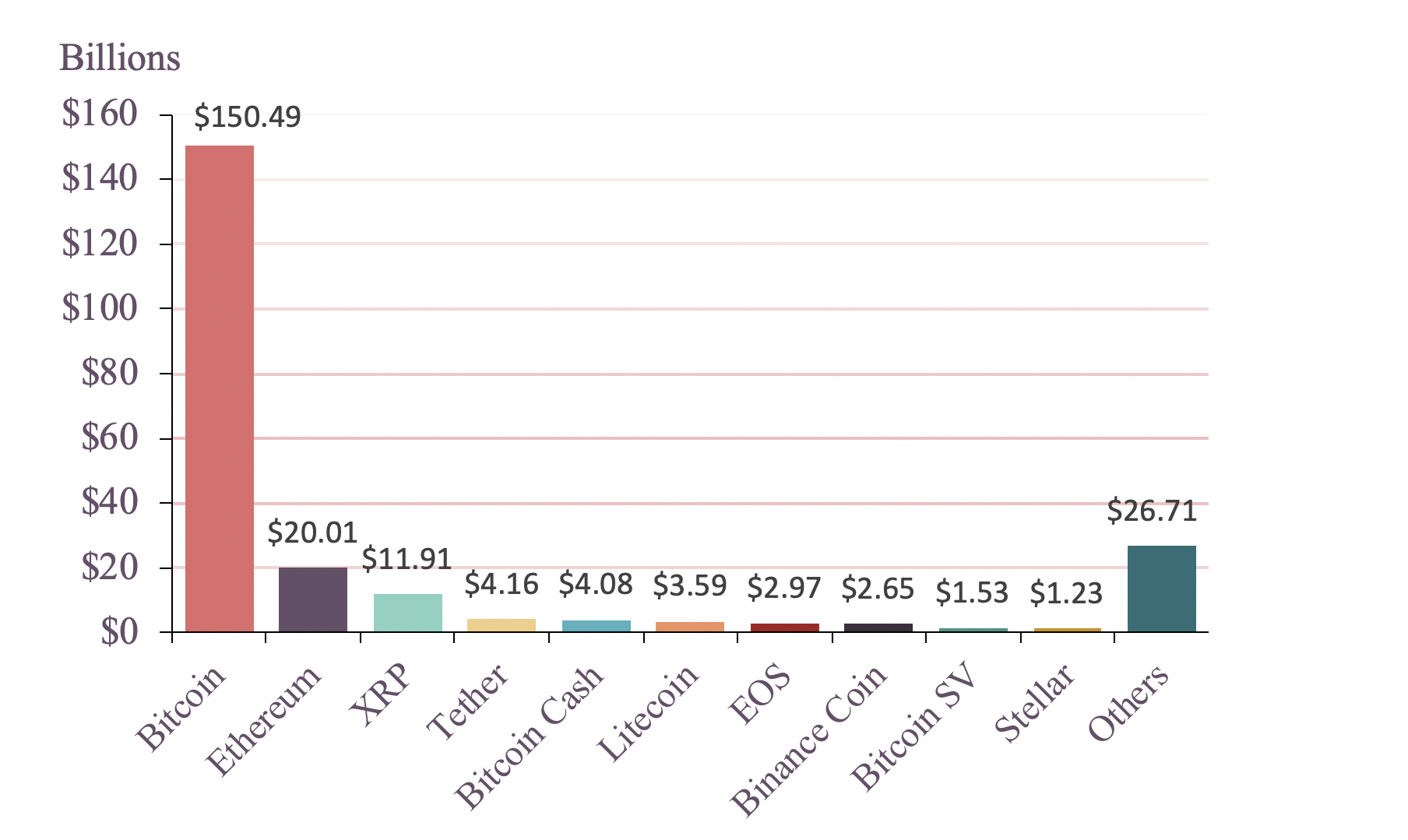}}
    \subfigure[]{\includegraphics[width=0.44\textwidth]{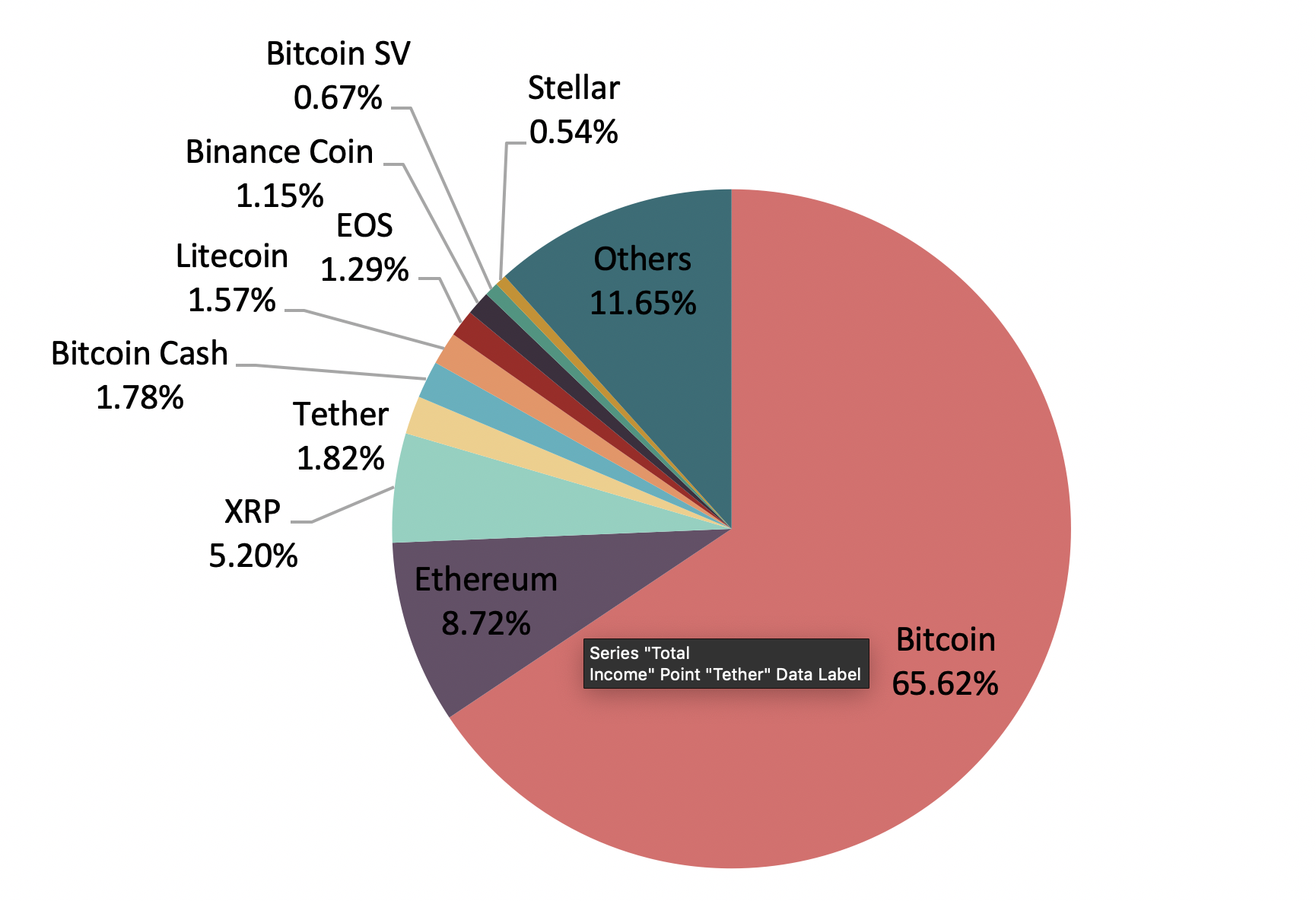}}
    \caption{The market cap distribution of 20539 cryptocurrencies (August 11, 2022, 11:00 AM UTC).} 
    \label{CryptoCurrency}
\end{figure*}

The first application of blockchain is the decentralized cryptocurrency, namely Bitcoin. Since then, thousands of alternatives to Bitcoin (or called altcoins) and other cryptocurrencies appear in the market. Fig.~\ref{CryptoCurrency} shows the market capitalization distribution of 20,539 cryptocurrencies as of August 11, 2022\footnote{https://coinmarketcap.com}. It's worth noting that many cryptocurrencies are not just employed for exchanging money but also have a wide range of other uses. In the following examples, we focus on their role as cryptocurrencies.

Litecoin~\cite{Litecoin} is a cryptocurrency that makes use of Scrypt encryption instead of Bitcoin's SHA-256 to make it resistant to ASIC mining. 
Dogecoin~\cite{Dogecoin} is based on Litecoin but has higher block rewards, an uncapped supply (leading to inflation), and lightweight blocks that only contain TXIDs. Litecoin and Dogecoin have block confirmation rates of 2.5 minutes and 1 minute, respectively. 
Namecoin~\cite{Namecoin} is the first cryptocurrency that is based on a fork of Bitcoin and introduces a decentralized naming service. It also brought into being the concept of \textit{merged mining}, which allows a miner to work on multiple blockchains at the same time. 
Peercoin is the first cryptocurrency employing a hybrid consensus algorithm that combines Proof-of-Work (PoW) and Proof-of-Stake (PoS), and Nxt is the first cryptocurrency to use a pure PoS consensus algorithm. 
In 2015, Ethereum introduced the Ether cryptocurrency, which is used to pay for ``gas'', or the computation needed to perform transactions and run smart contracts. 
Binance~\cite{binance}, a cryptocurrency exchange platform, was founded in 2017 and introduced the Binance Coin (BNB) cryptocurrency. In 2018, the EOS cryptocurrency was created by block.one and has been used on the EOSIO platform.

Non-fungible tokens (NFTs) are digital ownership records that are stored on a blockchain. They are similar to conventional proof-of-purchase documents, such as paper invoices or electronic receipts. 
NFTs can trace ownership information, which also ensures the authenticity and rarity of digital works~\cite{bhujel2022survey}. 
The first project similar to NFT is Colored Coin~\cite{ColoredCoins}, which was created in 2012 and could represent various assets such as property, coupons, and company shares. After the development of several years, the market potential of NFTs began to show and experienced a breakthrough in 2021.
According to Dappradar, the volume of NFT trades in Q3 2021 was almost \$10.7 billion, a 704\% increase from the previous quarter. The volume of NFT trades for the entire year of 2021 exceeded \$23 billion~\cite{dappradar}. There are currently many NFT trading platforms, including OpenSea~\cite{OpenSea}, SuperRare~\cite{SuperRare}, and Foundation~\cite{nftfoundation}, with more  emerging as time goes.

\subsubsection{Web 3.0 and Metaverse}

Web 3.0 is a decentralized web that is not controlled by any platform owner, but instead empowers content creators by giving them the ownership of their work and the corresponding rewards. This is made possible by utilizing blockchain as the foundation for Web 3.0's token economy. The idea of a metaverse, which seeks to merge the real and virtual worlds, has attracted a lot of interest, but security and privacy concerns present significant obstacles to its realization. Metaverses utilize various techniques, each with its own vulnerabilities, which make them vulnerable to various attacks. However, blockchain has emerged as a crucial component of metaverses, offering tamper-resistant decentralized ledgers that enable transparent and trustworthy computing environments. Although blockchain is currently used for DeFi and NFTs in metaverses, its full potential has yet to be realized~\cite{xu2022trustless}.  


\subsubsection{Trusted Computation, Storage and Access Control}

Trusted Computing is a concept that was first introduced by the Trusted Computing Group (TCG) to address computer security problems through hardware enhancements and associated software modifications. Various similar concepts have been proposed, including trusted systems and trustworthy systems. A trusted blockchain-enabled computing environment can provide trusted computation, storage, and access control services to all users, with the aid of cryptographic primitives such as Multi-Party Computation (MPC), Functional Encryption (FE), Verifiable Computing (VC), Homomorphic Encryption (HE), and Verifiable Storage (VS).
 
Trusted computation is a popular and significant area in the field of blockchain technologies. The work in~\cite{Enigma} makes use of secure multi-party computation along with a verifiable secret-sharing scheme to share and compute user data without compromising privacy. The scheme in~\cite{raman2018trusted} combines a lossy compression scheme with MPC to achieve scalability in blockchain-enabled secure computing. Andrychowicz \textit{et al.}~\cite{andrychowicz2014secure} specifically designed an MPC scheme to be used on the Bitcoin blockchain, along with Bitcoin transaction scripts to emulate a smart-contract-like secure logic on Bitcoin, which can be used for secure online gambling or decentralized autonomous organizations (DAOs). Federated learning and decentralized learning employ blockchain technologies to ensure the correctness and accuracy of the trained models when nodes may be faulty~\cite{xu2022spdl, wang2022incentive}.

In addition to trusted computations, secure storage in the context of blockchain technologies is another popular topic. Filecoin is a decentralized storage platform that employs its own cryptocurrency to facilitate the buying and selling of storage spaces. The Filecoin network is built on top of the InterPlanetary File System (IPFS), a peer-to-peer protocol for file sharing and storage. Filecoin aims to provide a more efficient and cost-effective alternative to traditional cloud storage services by allowing users to access unused storage capacity on devices around the world. FileDAG~\cite{guo2022filedag} is a decentralized storage network that takes a blockchain-based DAG for file-level deduplication and efficient updates. It stores only changes made to a file, if the file exists in the system, and employs a two-layer DAG-based blockchain for flexible and storage-saving file indexing. FileDAG outperforms other DSNs in terms of storage cost and latency. Zyskind \textit{et al.}~\cite{zyskind2015decentralizing} combined blockchain and off-blockchain storage to build a personal data management platform that focuses on preserving user privacy. They also added an MPC-enabled protocol for secure post-storage processing. Kishigami \textit{et al.}~\cite{kishigami2015blockchain} studied a blockchain-enabled digital rights management (DRM) scheme for secure digital content distribution. Dubovitskaya \textit{et al.}~\cite{dubovitskaya2017secure} proposed a blockchain-enabled system for securely and easily sharing patient electronic medical records (EMRs).

Current access control schemes for IoT devices are mainly based on authorization, and the majority lack accountability for access activities. This implies that once an access is granted, it is technically unlimited for a wide time span, with no restrictions. Such a mechanism grants an unlimited access privilege to the device's access server in stead of its owner by the device manufacturer, resulting in growing concerns in recent years. 
To address this issue, Tokoin~\cite{tokoin} instantiates the access power into accountable cryptographic assets. This shifts the access control scheme from an unlimited, authorization-based approach to an accountable, activity-based one, allowing device owners to have confidence that only the desired access takes place, and all access activities are audited.

\subsubsection{Internet of Things}

Blockchain technologies have the potential to improve the Internet of Things (IoT) in two ways. First, the decentralized and distributed nature of blockchain networks makes them well-suited for IoT. Currently, most IoT networks are centralized, which makes them vulnerable to single points of failures and can lead to increased communication latency as the number of devices grows. By contrast, a decentralized and distributed network topology could lead to a more efficient, automated, and self-governed IoT network. Second, blockchain technologies can address a number of security and privacy issues that remain unsolved using traditional methods. Its properties of immutability, traceability, and audibility make it well-suited for improving security and privacy of IoT networks. However, integrating IoT and blockchain is challenging due to the resource constraints of IoT devices, the complexity of managing heterogeneous devices and big data, and the scalability of blockchain systems. Below are some notable solutions to these challenges. 

Dorri \textit{et al.}~\cite{dorri2017towards} proposed an optimized hierarchical architecture that combines a centralized private Immutable Ledger (IL) with a decentralized blockchain to reduce overhead and increase trust. This architecture was implemented in the context of a smart home. 
Huh \textit{et al.}~\cite{huh2017managing} connected IoT devices to an Ethereum network for managing the devices. Novo~\cite{novo2018blockchain} addressed the issue of access management for IoT devices by having them interact with an Ethereum network through a management hub. 
Curb~\cite{xu2022curb} is a group-based SDN control plane that smoothly integrates blockchain and BFT consensus on edge, and supports dependable flow rule updates and adaptable controller reassignment.
Liu \textit{et al.}~\cite{liu2017blockchain} used a blockchain network to address the problem of verifying the integrity of IoT data. 
Ouaddah \textit{et al.}~\cite{ouaddah2017towards} proposed a theoretical framework for privacy-preserving access control in IoT networks. 
DistBlockNet~\cite{sharma2017distblocknet} is an IoT architecture that combines Software Defined Networking (SDN) and blockchain technology to provide protection and mitigate attacks. 
For more solutions, we recommend surveys such as \cite{ali2018applications}, which covers the use of blockchain in IoT, \cite{reyna2018blockchain} and \cite{panarello2018blockchain}, which focus on the integration of blockchain and IoT, and \cite{khan2018iot} and \cite{dorri2017blockchain}, which address IoT security and privacy. 

\subsubsection{Supply Chain}

Blockchain was originally developed to create a decentralized cryptocurrency called Bitcoin. However, it has many features that make it well-suited for creating a secure and distributed ledger of digital assets, which has led to an interest in adopting it to improve various industries beyond cryptocurrency. The supply chain industry is one that has the potential to benefit from blockchain technologies in many ways, particularly in improving trust through decentralization, traceability, transparency, and immutability. There have been a number of studies on using blockchain in supply chain, which have focused on both general problems and specific questions within the supply chain.

The study in~\cite{kshetri20181} systematically examines the ways in which blockchain can be used in supply chain and how it affects supply chain management objectives such as cost, quality, speed, dependability, risk reduction, sustainability, and flexibility. Kim and Laskowski~\cite{kim2018toward} proposed an ontology-driven approach to modeling the use of blockchain in supply chain. This approach employs both informal ontologies to improve blockchain development and business practices, and formal ontologies to aid in the creation of smart contracts. Korpela \textit{et al.}~\cite{korpela2017digital} investigated the process and data integration with blockchain technologies in supply chain. Saberi \textit{et al.}~\cite{saberi2019blockchain} discussed how blockchain can enhance the sustainability (economic, environmental, and social performance) of supply chain. Hackius \textit{et al.}~\cite{hackius2017blockchain} conducted a survey on the benefits of blockchain technologies for both supply chain management and logistics industries.

There are several industries in which their supply chains have the potential to be transformed by blockchain technologies, including manufacturing, agriculture, food, and shipping. For example, the potential benefits of using blockchain in the manufacturing supply chain have been analyzed in~\cite{abeyratne2016blockchain}, and a scheme for using blockchain in the agri-food supply chain was proposed in~\cite{tian2016agri}. Companies such as IBM, Deloitte, Corda, and Consensys have been working on transforming traditional supply chains with blockchain technologies.

\section{Challenges and Future Directions}
\label{sec:challenge}

\subsubsection{Scalability and Storage Cost} Blockchain's scalability is a crucial and ongoing subject of research and discourse within industry. Despite the numerous efforts being made to enhance the scalability of blockchain systems, it is likely that this topic will remain a focus as blockchain technologies continue to progress and gain wider adoption. To address the scalability challenges of blockchain technologies, a number of approaches have been explored, including layer-1 and layer-2 solutions. The DAG-based Byzantine Fault Tolerance (BFT) consensus, as a layer-1 solution, is expected to scale blockchain by separating the network communication layer from the consensus logic. Lots of asynchronous BFT consensus protocols appear in recent progresses, and DAG-based consensus has the best performance till now. Cross-chain interoperability is still a problem since we cannot efficiently operate on multiple blockchains yet without relying on centralized third parties. 
Besides, each of these approaches comes with its own advantages and disadvantages, and it is uncertain that, if any, will be effective in allowing blockchain systems to scale to the level suitable for widespread adoption.

Beyond scalability, users also have concerns about the storage cost of blockchain. The storage cost of an Ethereum full node is approaching 1 TB, which is a burden for a personal computer and obviously is unaffordable for mobile devices. With a growing storage cost, the number of full nodes will decrease, which harms the robustness of blockchain networks. In particular, the state data (e.g., world state tree) occupies a large amount of storage, which calls for novel designs of blockchain storage, leading to new approaches such as the stateless blockchain~\cite{boneh2019batching}. 

\subsubsection{Modular Architecture} As the blockchain industry continues to grow and diversify in terms of applications, users often rely on separate and distinct blockchain systems with varying functions. These systems often have data barriers that prevent the free flow of information, value transfer, and collaborative operations. Such barriers inhibit blockchain technologies' ability to fully utilize their core strengths of providing consensus and trust. Cross-chain technology, which aims to establish connections among blockchains by building a bridge between isolated systems, is a crucial technique for addressing this issue. However, current cross-chain solutions are insufficient to meet these needs. Modularization of blockchain technology offers a promising solution to this problem. A modular blockchain has the ability to adapt to different scenarios by providing variable support for networking, consensus, ledger, and other features. By doing so, the blockchain community can be united and barriers between blockchains can be broken down. 

\subsubsection{Security Concerns} Blockchain and smart contracts have the potential to provide a high level of security and privacy since lots of works are born with formal security proofs and open-sourced codes. However, like any complex system, they also introduce new security and privacy risks. One of the main security problems associated with blockchain is the possibility of hacking or tampering with the data stored on chain. This can occur through the use of malware or other cyber attacks, and can potentially compromise the integrity of the data stored on the chain. Smart contracts can also introduce security risks. If the code of a smart contract contains errors or vulnerabilities, it can potentially be exploited by attackers. There are  works investigating the vulnerabilities of smart contracts using traditional approaches such as fuzzing and symbolic execution. These methods do help find vulnerabilities and provide countermeasures. However, traditional approaches fall short of finding more in-depth vulnerabilities. With the development of NFT and Web 3.0 applications, smart contracts are becoming more complicated, resulting in hard-to-find and zero-day security issues. 

Even though many blockchains have strong security guarantee in design, we still have serious concerns about those centralized cryptocurrency exchanges, which might be uncontrollable. In November 2022, the cryptocurrency exchange FTX faced serious issues related to leverage and solvency that were reported by CoinDesk. This caused FTX to fail and had a significant impact on the cryptocurrency market, which saw billions of dollars in value lost and a market dropped below the valuation of \$1 trillion. The cryptocurrency industry has had a history of trying to prove to regulators, investors, and the general public that it is secure and reliable. However, the collapse of cryptocurrency exchanges can destroy confidence and harm the blockchain community. 

\subsubsection{Privacy Protection} In terms of privacy, the use of blockchain and smart contracts can potentially enable the tracking and recording of transactions in a way that is more transparent and traceable than traditional methods. This can be beneficial in certain situations, but it can also raise privacy concerns, particularly in cases where sensitive or personal information is involved. 
Traditional encryption algorithms such as AES and RSA~\cite{mahajan2013study}, are not sufficient to address privacy issues because verifiers/miners still need to validate the blockchain data. Recent blockchain projects, e.g., zkEVM~\cite{zkevm} and Monero, have implemented privacy-enhancing technologies such as zero-knowledge proofs and ring signatures to protect the privacy of users and their transactions, but they also come with issues such as low efficiency, high latency, and high energy consumption in terms of computation and/or communication overheads. 
In addition, in more complicated situations such as cross-chain, the privacy challenges discussed earlier would be even harder to address. Specifically, in cross-chain systems, the frequent interactions between multiple chains would exacerbate privacy concerns. In the context of different chains interacting with each other, it would be crucial to not only protect the privacy of multiple identities, data, and consensus but also expand the boundaries of privacy protection.

\subsubsection{Artificial Intelligence and Blockchain}
 Recently, AI models have surpassed humans in several tasks that were once considered impossible. For instance, DeepMind's AlphaGo defeated Jie Ke, the globally top-ranked player, in 2017, and the self-taught AlphaGo Zero achieved a 100-0 victory against AlphaGo in 2019. AlphaFold can predict a protein's 3D structure from its amino acid sequence. Besides, OpenAI's ChatGPT can provide detailed and articulate responses on a broad range of topics. AI-Generated Content (AIGC) has also been used to create artworks that resemble those produced by human artists. 
 
The integration of AI and blockchain technologies has the potential to revolutionize various industries. AI can increase productivity, while blockchain can improve production relationships. However, the fusion of these technologies is not yet fully developed, and related systems are not yet established. Despite this, their future synergy can have a significant impact.
On one hand, the implementation of AI technologies requires trustworthy regulators that are not controlled by third parties. Blockchain, with its decentralized and secure architecture, is currently one of the best options to provide the necessary regulatory framework.
On the other hand, blockchain can benefit from the introduction of AI to achieve intelligentization. This can lead to improved data management and analysis, as well as more efficient decision-making processes. The integration of AI and blockchain is a promising area for future development. As both technologies continue to evolve, it is important to explore their potential synergy and work towards establishing related systems that can unlock their full power.

\subsubsection{Applications} While there are many potential uses of blockchain technologies, there exist relatively few successful, large-scale deployments beyond finance. This can make it challenging for organizations to see the value and benefits of implementing a blockchain solution. There are a number of factors that have contributed to the slow adoption of blockchain technologies in industries beyond finance. 

First of all, blockchain technologies can be complex, particularly for those individuals and organizations without sufficient technical background. This can make it challenging for them to implement and integrate blockchain into their existing systems. Implementing a blockchain solution may require a significant investment in terms of resources and time, which can be a barrier for some organizations. 
Second, as with any new technology, there can be resistance to adopting blockchain solutions within organizations due to concerns about changing existing systems and processes, especially when blockchain is wrongly regarded as not essential. Moreover, blockchains introduce extra storage cost and latency, thereby the performance of the systems may be negatively impacted by a blockchain. 
Third, the regulatory environment surrounding blockchain technologies is still evolving, and this can create uncertainty for organizations that are considering blockchain. Besides, some organizations may be hesitant to adopt blockchain due to concerns about privacy and security of their data. This requires them to carefully evaluate the security and privacy risks associated with the adoption of blockchain and take appropriate measures to mitigate the corresponding risks.
Fourth, expanding the use cases of blockchain technologies is not trivial. It is challenging for small companies to develop permissionless blockchains, as they are complex and difficult to manage. As a result, many companies are attempting to create permissioned blockchains, which rely on the assumption that the consensus nodes can be trusted. However, permissioned blockchains, unlike permissionless ones, do not effectively address the issues that arise when users do not trust each other, thereby failing to experience the full benefits brought by blockchain.

\section{Conclusion}
\label{conclusion}
This article presents a thorough analysis on the layer-1 and layer-2 scaling solutions of blockchain from a macro perspective, and proposes a modular blockchain analytic framework with which a blockchain system can be scrutinized from a micro perspective. Particularly, the three essential components of a blockchain system, namely the network, consensus protocol, and distributed ledger, are comprehensively surveyed and analyzed. Major techniques that can be integrated with blockchain to enhance its performance, as well as the emerging applications that rely on blockchain to offer specific services, are then articulated. The future directions of blockchain development and the corresponding challenges are finally examined. The goal of this article is to deepen our comprehension on the capabilities and limitations of the current blockchain technologies and to provide insights into its potential future developments.

\section{Acknowledgements}
We are thankful to Arkady Yerukhimovich for many deep discussions.

\bibliographystyle{ACM-Reference-Format}
\bibliography{References}

\end{document}